\documentclass[fleqn,usenatbib]{mnras}

\usepackage{newtxtext,newtxmath}
\usepackage[T1]{fontenc}

\DeclareRobustCommand{\VAN}[3]{#2}
\let\VANthebibliography\thebibliography
\def\thebibliography{\DeclareRobustCommand{\VAN}[3]{##3}\VANthebibliography}

\usepackage{graphicx}	% Including figure files
\usepackage{amsmath}	% Advanced maths commands
\usepackage{subcaption}
\title[Plasma chemistry in L/T dwarfs]{Plasma chemistry and electron-moderated pathways in substellar atmospheres: a new perspective on the L/T transition}

\author[M. I. Swayne et al.]{
Matthew I. Swayne$^{1}$\thanks{E-mail: Matthew.Swayne@glasgow.ac.uk}, Declan A. Diver$^{1}$, Craig R. Stark$^{1}$\thanks{E-mail: craig.stark@glasgow.ac.uk}
\\
$^{1}$SUPA, School of Physics and Astronomy, Kelvin Building, University of Glasgow, Glasgow, G12 8QQ, Scotland, UK
}

\date{Accepted XXX. Received YYY; in original form ZZZ}

\pubyear{\the\year{}}

\begin{document}
\label{firstpage}
\pagerange{\pageref{firstpage}--\pageref{lastpage}}
\maketitle

% Abstract of the paper
\begin{abstract}
The long‑standing puzzle of the CO/CH$_{4}$ transition in brown dwarfs endures.% because conventional chemistry lacks a crucial driver. 
  ~Although the bulk spectral evolution across an atmosphere can be accounted for through thermal equilibrium cloud models, the behaviour in the NIR remains unaccounted for, indicating that additional, non-thermal processes may influence atmospheric chemistry alongside conventional pathways.
We explore cloud‑driven electrical activation, where low‑energy sparks to full lightning discharges, unlocks non‑equilibrium reaction pathways inaccessible under thermal conditions alone. To quantify this, the aim of this paper is to model the electron-moderated atmospheric chemistry with \texttt{SPARCKS}, a bespoke zero‑dimensional code that solves the coupled set of particle balance equations for substellar plasma activation and reaction kinetics, focusing on the key CO–CH$_{4}$ electron-moderated chemistry across the parameter ranges $T_{\rm gas} \in [700, 1600]\,\mathrm{K}$ and $T_{e} \in [2, 5]\,\mathrm{eV}$. We simulate a 1 microsecond pulse, representing a short dart-stepped leader; and, two pulsed systems with $(t_{\rm on}, t_{\rm off}) = (10^{-8}~\mathrm{s}, 10^{-6}~\mathrm{s})$ and $(10^{-9}~\mathrm{s}, 10^{-9}~\mathrm{s})$, representing small‑scale inter‑grain discharges, consistent with a typical characteristic substellar atmosphere. Our results show that even modest, physically plausible energies can strongly perturb atmospheric composition: an electron energy of 3.0 eV is sufficient to halve the CH$_{4}$/CO ratio in our sample atmosphere within one microsecond. Beyond the CO–CH$_{4}$ system, electron‑moderated plasma chemistry exerts a far‑reaching influence on substellar atmospheric composition.
\end{abstract}

% Select between one and six entries from the list of approved keywords.
% Don't make up new ones.

%%%%%%%%%%%%%%%%%%%%%%%%%%%%%%%%%%%%%%%%%%%%%%%%%%

%%%%%%%%%%%%%%%%% BODY OF PAPER %%%%%%%%%%%%%%%%%%

\section{Introduction}

Brown dwarfs (BDs) straddle the L/T transition, a regime where conventional equilibrium chemistry breaks down. Thermal equilibrium models across the L/T transition can reproduce much of the observed spectral behaviour of brown dwarfs, particularly when combining cloudy and cloudless atmospheres \citep[e.g.][]{2015Marley,2021Marley,2024Morley,2024Biller}. These models account for CO in L-dwarf atmospheres without invoking additional processes, and retrieval studies of L-dwarfs indicate that such equilibrium frameworks provide a good first-order description of the atmospheric chemistry \citep[e.g.][]{2017Ben} However, systematic residuals and clear signatures of disequilibrium chemistry persist, especially in T dwarfs where CO is observed in excess of equilibrium expectations, explored in \cite{2024Muk}. Photochemistry \citep{2014Moses} and vertical mixing \citep{2003Saumon} have been invoked to explain rapid shifts in CO and CH$_{4}$ abundances, but as of yet does not fully cover the dramatic shift to bluer colour and accompanying J-band brightening in the NIR \citep{2012Morley}. This raises the possibility that non-equilibrium chemistry could have a crucial contribution. 
We propose that electrical activation, lightning and plasma discharges, drives a suite of non-equilibrium reactions otherwise inaccessible thermodynamically. Even transient plasma events can reshape atmospheric chemistry, and could be caused by the dust collisions in brown dwarf atmospheres \citep{2011Helling,2013helling,2013Stark,2016Hodosan}, generating low-energy sparking to full lightning discharges.
This could then fit within the physical explanation for the observed CO/CH$_{4}$ transition from L- to T dwarfs, the original delineator between L- and T-dwarfs \citep{1999Kirkpatrick}. 
Electrical activation is not proposed as the driver of the L/T transition, but as a contributor to the observed disequilibrium chemistry that coexists with the thermal‑equilibrium baseline.

The L/T transition is a key evolutionary phase in BD dusty atmospheres, marked by near-infrared colours shifting dramatically from very red in late-L and early-T dwarfs to very blue in mid-T dwarfs \citep{2004Knapp}. This change occurs over a narrow effective temperature range of approximately 1400–1200 K and is thought to reflect the disappearance of dust clouds, either through dispersal or by sinking below the photosphere \citep{2006Burrows} either as the L-dwarf cools towards the T-dwarf phase, or in comparing an object formed as a L-dwarf to one formed as a T-dwarf.
We define clouds in this paper as the large-scale atmospheric structures composed of ensembles of dust grains, individual solid particles that form through condensation processes.
Spectrophotometric variability across the transition suggests that cloud coverage is spatially inhomogeneous \citep{2013Apai}, reinforcing the view that atmospheres evolve from cloudy to “cloud-free.”

In BD atmospheres, clouds play a critical role in disequilibrium: dust charging and discharging is unavoidable and studied in a variety of planetary contexts \citep{lacks2007effect,krauss2003experimental,zheng2003laboratory}.
Electrical activation occurs at the level of the dust grains themselves, where charge is generated through triboelectric interactions, grain–grain collisions, and differential settling; creating transient activation that seeds departures from conventional thermodynamics \citep{2014Stark} and is of a higher energy than those that would be caused by sources of thermal ionization expected to be present in brown dwarfs \citep{2015Rod}.
These grain‑scale charging processes can then lead to either small‑scale inter‑grain discharges or larger‑scale discharge events when charge separation develops across extended cloud regions.

Electrostatic discharges are expected to open new chemical routes in the presence of H$_2$O, CO$_2$, and CH$_4$. For example, electron-moderated reactions could destroy CH$_4$ ($e^{-} + ~$CH$_{4} \rightarrow $\:CH$_{3}$ + H$^{-}$) or produce CO ($e^{-} + $\:CO$_{2}\rightarrow $\:CO\:$ + $\:O$^{-}$). Such discharges can plausibly generate vibrationally excited species that lower activation barriers, enabling reactions like steam reforming (CH$_{4}$ + H$_{2}$O$\:\rightarrow\:$CO$ \:+\: $3H$_{2}$) and dry reforming (CH$_{4} \:+\: $CO$_{2} \rightarrow\: $2CO $+$ 2H$_{2}$) at rates far exceeding equilibrium expectations.

Sub-stellar atmospheres from L- to Y- dwarfs can exhibit some of a variety of different emission sources including radio \citep{2017Williams}, X-ray \citep{2010Berger}, and optical H-alpha emission \citep{2016Pineda}, clear signatures of plasma activity.
Observed due to large-scale magnetospheric systems, the resultant auroral emission in particular is evidence that a stable population of electrons can be maintained upon a substellar dwarf \citep{2015Hallinan,2017Pineda,2023Kao}.
Electrical activity can affect the formation of certain molecules, resulting in the weakening of spectral absorption bands such as the 2.7 ${\rm \mu m}$ water line \citep{2014Sorahana} and the increase in the abundance of small hydrocarbon molecules like CH and CH$_{2}$ \citep{2014Bailey}.
Brown dwarf lightning events are therefore likely to create a complex mixture of metastables, radicals, vibrationally excited species, free electrons, and ions -- lowering energy barriers and triggering chemical pathways that would otherwise be energetically inaccessible.

Crucially, this physically sits within what would be expected for the CO/CH$_{4}$ transition. In L-dwarfs, abundant clouds sustain charging and discharging, driving plasma chemistry that preferentially destroys CH$_{4}$ and produces CO. As the atmosphere cools over its lifetime and evolves toward the T-dwarf regime, clouds dissipate, electrical activation is quenched, and plasma pathways vanish. Without this activated chemistry, CH$_{4}$ can re-emerge as the dominant species due to conventional equilibrium chemistry while CO declines.
Although individual brown dwarfs cool from L  to T type conditions over time, our model is agnostic to specific evolutionary tracks. Instead, our model describes how electron moderated plasma chemistry behaves under the atmospheric conditions characteristic of each spectral type. Therefore, our modelling framework can be viewed either as a description of the contrasting properties of L  and T dwarf atmospheres or as the sequence of chemical behaviour an individual object would experience as it cools with time.

This work sits within the rapidly evolving landscape of substellar science. Since the 2000s, the field has been defined by the discoveries of brown dwarfs \citep{1995Nakajima} and exoplanets \citep{1992Wolszczan}.
Through multiple methodologies, most commonly the radial velocity and transit methods, these objects have been found in exponentially increasing quantities.
As instrumentation, computation and data analysis techniques have improved, so has the complexity of the science that can be performed observing or simulating these extrasolar objects.

A field that typifies this is that of substellar atmospheres.
It has evolved from the first tentative detections of molecules by \textit{Hubble} \citep{2002Charbonneau}, to full atmospheric retrievals from analysing spectra of high resolution and precision.
With JWST and other upcoming ground- and space-based missions and observatories, we are now in an era of unprecedented high precision data from spectroscopy to photometry.
As observed data increases in complexity so must the atmospheric models we use to interpret them.
For simplicity, past retrievals of substellar atmospheric parameters have mostly used models where atmospheres are held to be in Local Theromodynamic Equilibrium (LTE).
However in reality we know this to not be the case, with multiple observations of our own solar system requiring or suggesting the occurrence of non-LTE processes in the Earth \citep{1971Charters}, Venus \citep{2009Gilli}, Mars \citep{2005Lopez}, Titan \citep{2011Garcia,2013Rezac} and both Jupiter and Saturn \citep{1999Drossart}, mainly through the spectral observations of excitation reactions achieved through solar pumping.
With the recent influx of high precision JWST data this has come to a head, with multiple examples of substellar observations where the current suite of atmospheric retrieval models are unable to fully characterise unusual or complex observations \citep[e.g.][]{2022Parviainen,2024Biller}.
Non-LTE processes are being considered to attempt to explain these discrepancies, with their inclusion in atmospheric models an ongoing goal for modellers to better interpret the data from current and upcoming missions.
Thus, our use of electrically-active chemistry to model the L-T transition follows the current re-examination of unsolved problems in the field, especially at the different transition regions of the substellar.

Observationally, Lightning discharges are one such source of non-LTE cloud processes and a driver of electrically-activated chemistry that has been observed in solar system bodies other than the Earth, most pertinently Jupiter \citep{2023Kolmasova}.
With Jupiter-like exoplanets and brown dwarfs sharing multiple behaviours it has been proposed that brown dwarfs will also host lightning processes \citep{2016Hellinga,2016Hellingb}.
Additionally, lightning has previously been proposed to enhance spectral features and be able to mask biomarkers in the atmosphere of a terrestrial planet \citep{2024Barth} and is proposed as a potential creator of primordial life based on the implications from the famous Miller-Urey experiment \citep{miller1959organic}.
As brown dwarf atmospheres are easier to characterise, being able to understand the commonality of lightning processes in the galaxy through observing its impact on these objects (as well as becoming familiar with how to detect this impact), would be relevant to the study of all exoplanetary atmospheres and the hunt for life.
If electrically-active chemistry is modelled to have a significant impact on substellar chemistry, it thus could not only help clear up debates on the cause of the L-T transition \citep{2019Tremblin,2019Vos}, but be of import to many exoplanetary objects as well.

This paper presents the first quantitative framework and analysis for examining the effects of electrically-activated chemistry upon key chemistry in a transition brown dwarf atmosphere, showing its effect on key CO-${\rm CH_4}$ chemistry.
We do this for a range of temperatures appropriate for these brown dwarfs, whose effective temperatures span a narrow band from $\sim$ 1400-1200 K, covering spectral types L8-T1 \citep{2019Vos}.
We first introduce our 0-D chemical model \texttt{SPARCKS} and detail our chosen atmospheric scenarios.
We then present the results of our simulations for two different discharge test-cases, representing the edges of a lightning strike and discharges between electrified grains.
Finally, we infer from our results the impact of this chemistry in transition dwarf atmospheres and its potential to influence the wider field.

\section{Electron-moderated chemistry in substellar atmospheres}
\label{sec:methods}

\subsection{Model Equations}
To model the plasma (electron-moderated) chemistry in a brown dwarf atmosphere we used a zero-dimensional, global model where the spatial variation is assumed in order to focus on the complex chemical pathways present in the system. This does not restrict the broader applicability of the model, nor does it preclude the plasma being spatially non-uniform -- analytical functions can be used to define any spatial variation. By neglecting spatial dependence, the governing system simplifies to a coupled set of ordinary differential equations -- particle balance (rate) equations. For each species represented in the model, an individual particle balance equation must be formulated to capture its rate of change under the influence of chemical reactions and system losses. In general, particle balance equations are expressed in the form:
\begin{equation}\label{rate0}
\frac{dn_{\alpha}}{dt}=S_{\alpha}-W_{\alpha}+\sum_{i=1}^{N_{\alpha,c}}a_{i}k_{i}n_{\alpha}n_c-\sum_{j=1}^{N_{\alpha,d}}a_{j}k_{j}n_{\alpha}n_{d},
\end{equation}
where $n_{\alpha}$ is the particle number density of species $\alpha$; $S_{\alpha}$ is the mass flow rate in and out of the vessel; $W_{\alpha}$ represents the losses to the material surfaces in the system, such as dust grains; $N_{\alpha,c/d}$ is the number of construction/destruction reactions involving species $\alpha$; $a_{i/j}$ is the number of particles gained/lost per collision; $k_{i/j}$ is the rate constant, and $n_{c/d}$ is the number density of the reactant constructively/destructively interacting with species $\alpha$. For a simplified substellar atmospheric system, we will assume that $W_{\alpha}=S_{\alpha}=0$ and for the plasma chemistry of interest the right-hand side of Eq. \ref{rate0}, becomes:
\begin{equation}\label{rate}
\frac{dn_{\alpha}}{dt} = \sum_{i=1}^{N_{\alpha,c}}a_{i}k_{i}n_{\alpha}n_c-\sum_{j=1}^{N_{\alpha,d}}a_{j}k_{j}n_{\alpha}n_{d}
\end{equation}

\texttt{SPARCKS} (Substellar Plasma-Activated Reaction Chemical Kinetics Solver) is a bespoke substellar plasma activation and reaction kinetics code that for a given set of electron-moderated chemical reactions, solves the coupled set of particle balance (rate) equations (Equation \ref{rate} for all $\alpha$), using a simple 4th order Runge-Kutta numerical scheme.
It uses this to determine the temporal evolution of an atmospheric system under the effect of electrical activation. Electrical activation can be the result of small-scale, inter-grain electrical discharges or large-scale, cloud lightning.
The contribution to the overall sum will be either positive for a synthesis reaction or negative to a decomposition reaction and through summing these the change in $n_\alpha$ for each species is calculated.
This process can then repeat for each step until the end of the simulation.

\subsection{Electron-moderated chemical reactions}
For the electron-moderated chemical reactions of interest \texttt{SPARCKS} uses rate constants and cross sections from multiple sources (given in Appendix \ref{sec:appendix}).
The majority of reactions for these added molecules were taken from the Supplement to \cite{wang2018}. 
To calculate rate constants for those reactions which were listed as cross-section dependent (mainly electron-neutral reactions such as those we are interested in), they used a Boltzmann solver inbuilt into the chemical kinetics model \texttt{ZDPlaskin} \citep{2008Panche}, \texttt{BOLSIG+} \citep{2005Hagelaar}.
For cross section-dependent reactions we replicated this approach, using \texttt{BOLSIG+} to calculate all electron-neutral reactions.
\texttt{BOLSIG+} seeks to provide steady-state solutions of the electron Boltzmann equation in a uniform electric field, described in detail in \cite{2005Hagelaar}.
The imposed electric field determines the steady state electron energy distribution function, which sets the rate coefficients obtained by averaging the cross sections over the electron energy distribution function. It can calculate solutions either over a range of electric fields or a range of energies, through calculating the electric field for the desired energy.
By running \texttt{BOLSIG+} over our electron temperature range, we can obtain a grid of cross-section dependent rate coefficients.
We calculate these assuming Maxwellian electron energy distribution functions, representing a non-thermal plasma where the electron temperature is much greater than the heavy particles (ions and neutrals) and that electron-electron collisions are sufficiently strong to define an electron temperature.

\subsection{Initial conditions}

We use \texttt{SPARCKS} to model over a range of gas and electron temperatures.
The gas temperature represents the temperature of the ambient gas in the atmosphere and is set from 750 to 1600 K in steps of 50 K, so as to sample over multiple possible atmospheric scenarios across the L-T transition.
The electron temperature represents the temperature that electrons are heated to during the discharge forming a non-LTE ionised plasma.
In the centre of lightning discharges this can reach tens of thousands of electron volts (eV) and will result in the mass dissociation and ionization of molecules.
We simulate a range of electron temperatures from 2.0 to 5.0 eV in steps of 0.05 eV.
This represents the lower electron temperatures in two scenarios of interest.
Firstly, the edge of lightning discharge events where ionised molecules and dissociated radicals will interact with material outside the path of the discharge.
Secondly, lower-energy inter-grain sparking between charged dust grains.
A lightning discharge also paints dust grains with electrons, leaving a population of charged grains behind.
Sparking then occurs when the grains pass close enough to trigger a small discharge event.
This process will then repeat with the next charged grain encounter resulting in multiple discharge events for the same pocket of gas.
In this way smaller discharge events over a continual timescale can enact significant change to an atmosphere, precluding a reliance on the frequency of large-scale lightning events.
Possibly one discharge could be all that is required for a persistent electrification of part of the atmosphere.

To examine the effect the discharge has on the transition-critical species ${\rm CH_4}$ and CO in particular, we simulate for a novel atmosphere 70 \% molecular Hydrogen (${\rm H_2}$). 
Water (${\rm H_2O}$), methane (${\rm CH_4}$) and carbon monoxide (${\rm CO}$) are each held to 10 \%; as they are all significantly present species in a 1500K L-dwarf according to the results of \cite{2013Stark}.
%This is a higher fraction of atmospheric content than we would expect from models but was chosen to highlight the effect on these species critical to the L-T transition.
We also have selected minority species according to \cite{2013Stark} (${\rm CO_2}$, H and C) which we set at $10^{-4}$ \% of the total combined number density.
The model outputs the number densities of CO and ${\rm CH_4}$ normalised to the total background gas density, and we analyse the discharge‑driven deviation in their ratio. Because the absolute abundances of CO and ${\rm CH_4}$ in the target environment are not well constrained, we adopt a representative configuration in which both species are present as minority constituents at equal initial abundances (10\% of the background). Our aim is not to predict their precise concentrations, but to examine how a discharge perturbs their relative abundances compared to the gas‑phase baseline. In this minority‑species regime, the evolution of CO and ${\rm CH_4}$ is governed primarily by discharge‑driven pathways — electron‑impact dissociation, radical production, and ion–molecule chemistry — which depend far more on the electron energy distribution than on the exact initial abundances or the background gas temperature. As a result, the qualitative deviation in the CO/${\rm CH_4}$ ratio is far less sensitive to these initial conditions than to the presence of the discharge and the structure of the reaction network itself. This single, physically plausible case therefore serves as an indicative demonstration of the type of chemical restructuring that a discharge can induce.

In a transition brown dwarf, the ratios of ${\rm CH_4}$, ${\rm H_2O}$ and CO with respect to the background number density would realistically be to the order of $10^{-5}$ - $10^{-3}$ as seen in \cite{2013Stark} and \cite{2014Zahnle} at the temperatures of the L-T transition.
These values provide a useful baseline when considering potential observational impacts. The observational significance of the electron-moderated plasma chemistry will depend on the underlying absolute abundances: a given fractional change has different radiative consequences if the background mixing ratio is $10^{-4}$ rather than 0.1. However, the aim of our model is not to predict the absolute abundances as a result of the plasma chemistry, but to identify the chemical pathways through which a discharge can alter the CO/${\rm CH_4}$ ratio. In this context, the key diagnostic is the relative deviation in the CO/${\rm CH_4}$ ratio, which is governed primarily by the electron‑moderated chemistry.
Furthermore, the influence of the plasma chemistry does not need to be limited to a single electrical discharge event. In a dynamically active dust cloud, small-scale inter-grain discharges can occur repeatedly between different grain pairs or between large-scale charge separated regions. As a result, such discharge events can sample fresh local gas regions while still operating within the same general atmospheric volume. The cumulative effect of multiple, spatially distributed discharges can process a non‑negligible fraction of the local gas, even when the background mixing ratios of CO and ${\rm CH_4}$ are well below 0.1. Therefore, the discharge‑driven restructuring of the CO/${\rm CH_4}$ ratio can remain chemically significant, while the translation into observable spectral signatures is a separate radiative‑transfer problem beyond the scope of the present study.

\subsection{Timescale considerations}
In substellar atmospheres, electrical activation can occur via small-scale inter-grain microdischarges or large-scale lightning discharges. To model the effect of a single high-energy lightning event we simulate a 1 microsecond pulse (pulse here defined as the length of time during which the model assumes the high electron temperature of the discharge), representing a short dart-stepped leader \citep{2023Kolmasova}. By modelling the change of atmospheric composition over this timescale, we represent the effect of the leader pulse's ionization front travelling through a point in the atmosphere. We set an ionization level (and thus normalised number density of electrons) at $10^{-6}$ based off models of dust grains in clouds, namely the dust grain to gas ratio, the number of electrons present on the grain surface and the amount of electrons shed in a discharge reaction \citep{2011Helling,2020Woitke}.
This represents the initial injection of electrons from the start of the discharge.
For comparison, the expected thermal degree of ionization for a typical L and T dwarf at high pressures of $10^{-1}$ to $10^0$ bar, characterised by $(T_{\mathrm{eff}},log(g))=(1800 K,3)$ and $(1000 K,3)$, is $f_{e}\approx10^{-7}-10^{-6}$ and $10^{-10}-10^{-8}$ respectively \citep{2015Rod}.

Inter-grain electrical discharges arise when charged dust grains approach one another closely enough for breakdown to occur, the process where the interposing air suddenly allows a current to be conducted between the grains. In a substellar dust cloud, the characteristic rate of such discharges is therefore expected to be comparable to the dust-dust collisional mean free time. The frequency of inter-grain discharges will therefore rely on the astrophysical quantities that govern collisional mean free time, such as dust-to-gas density, dust grain volume/surface area and the gas effective temperature. The timescales of this behaviour are analogous to the intermittency seen in pulsed plasma systems used in technological applications, and thus allows us to use the chemical pathways and reactions rates of plasma physics for our chemical modelling.

In pulsed plasma systems, the temporal modulation of the discharge can be exploited to selectively enhance the formation of favourable chemical species. During the on-pulse, the timescale can be tuned to match the characteristic creation time of a target species, ensuring its growth through the desired reaction pathway. For a reaction $dn_A/dt = k_1 n_A n_B$, the synthesis timescale is $t_c = (k_1 n_B)^{-1}$. As long as $t_{\mathrm{on}} \geq t_c$, species $A$ will accumulate. Conversely, during the off-pulse, the timescale can be adjusted to suppress destruction pathways. For a loss process $dn_A/dt = -k_2 n_A n_C$, the decomposition timescale is $t_d = (k_2 n_C)^{-1}$. If $t_{\mathrm{off}} < t_d$, species $A$ does not have sufficient time to decay, thereby enhancing its net production. This interplay between on- and off-pulse durations provides a controllable mechanism for steering plasma chemistry toward desired outcomes. To enhance the destruction of CH$_{4}$ in favour of CO, a similar temporal modulation could be applied. The key CH$_{4}$ decomposition reaction is CH$_{4}+e^{-}\rightarrow$~CH$_{3}+~$H$~+$~e$^{-}$ which has a characteristic timescale of $t_{0}\approx 10^{-9}$~s, so as long $t_{on}\gtrsim t_{0}$ then CH$_{4}$ will be optimally destroyed.

In a substellar atmosphere the on-pulse duration will dictated by the timescale for the injection of the electron population into the atmosphere. This timescale can be estimated by considering the timescale of evolution of an electron avalanche or streamer discharge. For a typical streamer propagating with speed $v\approx 10^{5}$~ms$^{-1}$, and with length $L\approx n_{d}^{-1/3}$, a streamer timescale is given by $t\approx n_{d}^{-1/3}/v\approx 10^{-9}-10^{-8}$~seconds for $n_{d}\approx10^{9}-10^{12}$~m$^{-3}$. The on-pulse timescales are consistent with inter-grain coronal discharges \citep{chauzy1989electric,sin2014corona}.  The off-pulse duration will be driven by the timescale for inter-grain interactions, which can be approximated by the dust grain collisional mean free time given by $t_{\rm coll}=(4\pi\sqrt{2}a^{2}v_{d}n_{d})^{-1}$, where $n_{d}$ is the dust particle number density, $a$ is the dust radius and $v_{d}$ is characteristic dust grain speed. For $a\approx10^{-5}$~m, $n_{d}\approx10^{12}$~m$^{-3}$ and $v_{d}\approx10^{2}$~ms$^{-1}$, $t_{\rm coll}\approx10^{-6}$~s. Due to the variation in the dust cloud variables, there will be a range of collision timescales. For example, for a dense substellar dust cloud with $n_{d}\approx 10^{13}$~m$^{-3}$ and a population of dust grains with $a\approx 10^{-5}$~m, we would expect $t_{\rm coll}\approx10^{-8}$~s. These timescales are consistent with \cite{2020Woitke}. 

For the models presented here, we will consider two pulsed systems encapsulating the relevant timescales: Scenario 1. $t_{\rm on}=10^{-8}$~s and $t_{\rm off}=10^{-6}$~s; and Scenario 2. $t_{\rm on}=10^{-9}$~s and $t_{\rm off}=10^{-9}$~s, consistent with a typical characteristic substellar atmosphere. Scenario 1 represents a less dense cloud with infrequent collision timescales, whereas Scenario 2 represents a more dense cloud with frequent collisions. For case 1., we simulate at our chosen electron temperatures for 10 ns before lowering it to the activation energy of the lowest excitation reaction, 0.16eV, for 990 ns. We repeat this four times to a total duration of 4 microseconds. In both cases, we split the electron number density so that the contribution of each pulse will add up to an introduced ionization level of $10^{-6}$ by the end of the simulation.

\begin{figure*}
    \centering
    \includegraphics[width=0.99\linewidth]{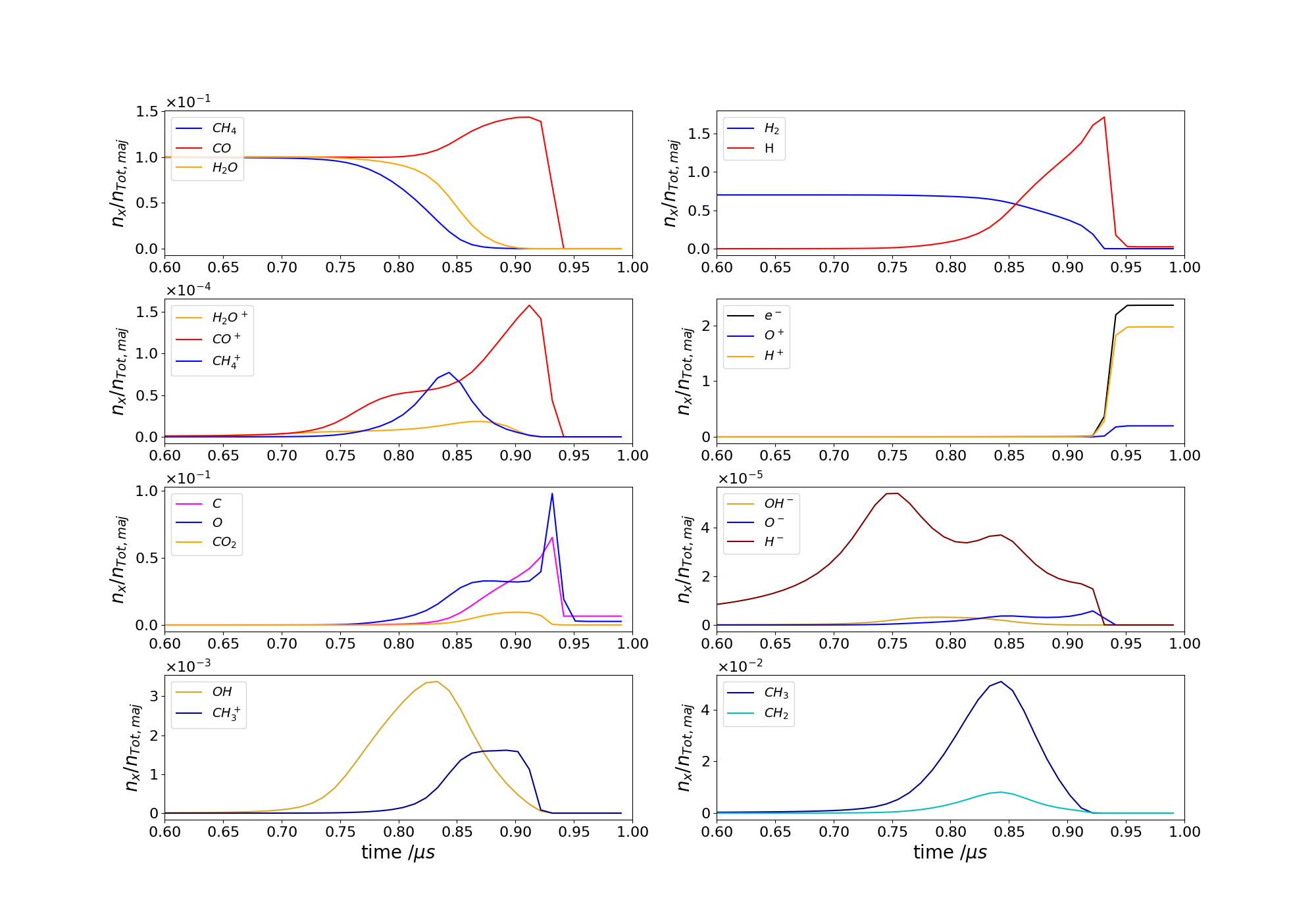}
        %\label{fig:majplotpls}
    \caption{The change in multiple different species in the single-pulse scenario at 3.05eV for a gas temperature of 1600K. Note that early‑time growth is not visible on the linear scale, so the plot begins once the signal becomes resolvable. Displayed on the y axis is the normalised number density of each species, where ${\rm n_x}$ is number density of species x and ${\rm n_{tot,maj}}$ is the combined number densities of ${\rm H_2, CH_4, H_2O}$ and CO. Our initial species show small-scale reactions until an increasing electron number density begins to cause electron-moderated dissociation and ionization to dominate. The onset of large-scale dissociation and ionization occurs within one microsecond for a 3.05eV discharge. At higher energies this behaviour will occur earlier whereas for small energies it would occur beyond the timescale of one microsecond.}
    \label{fig:mplotsingtg}
\end{figure*}

\begin{figure*}
    \centering
    \includegraphics[width=0.99\linewidth]{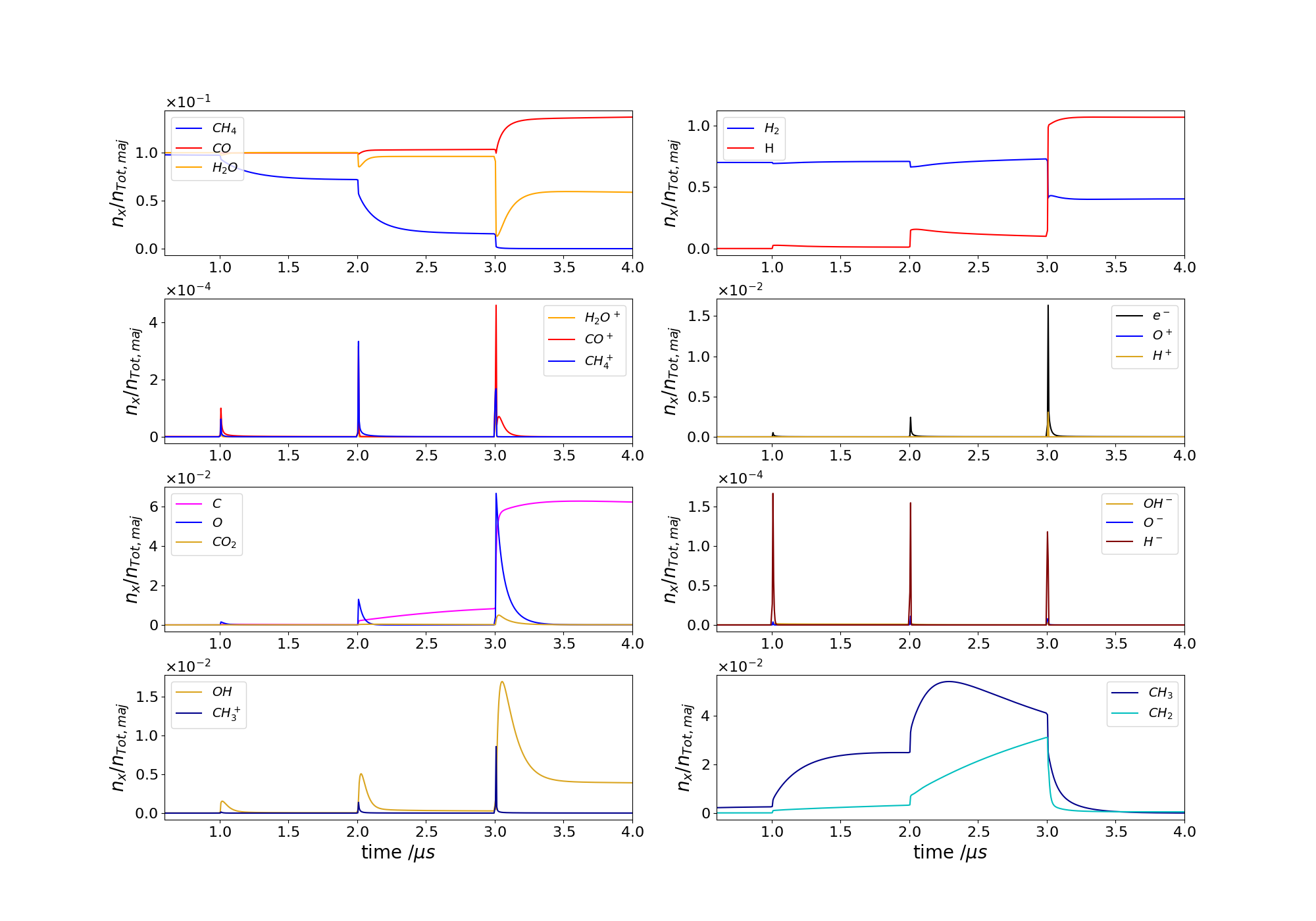}
    \caption{The change in multiple different species in the first multi-pulse scenario (10ns on-pulse, 990ns off-pulse) at 4.0eV for a gas temperature of 1600K. Note that early‑time growth is not visible on the linear scale, so the plot begins once the signal becomes resolvable. Displayed on the y axis is the normalised number density of each species, where ${\rm n_x}$ is number density of species x and ${\rm n_{tot,maj}}$ is the combined number densities of ${\rm H_2, CH_4, H_2O}$ and CO. The overall trends of destruction and construction are similar to the singe-pulse scenario, but the shorter time at high electron energies result in small spikes in electron-associated reactions followed by periods of recombination. Therefore a higher electron temperature is required to see the major dissociation of the initial species in the simulation.}
    \label{fig:mplotpls1}
\end{figure*}

\begin{figure*}
    \centering
    \includegraphics[width=0.99\linewidth]{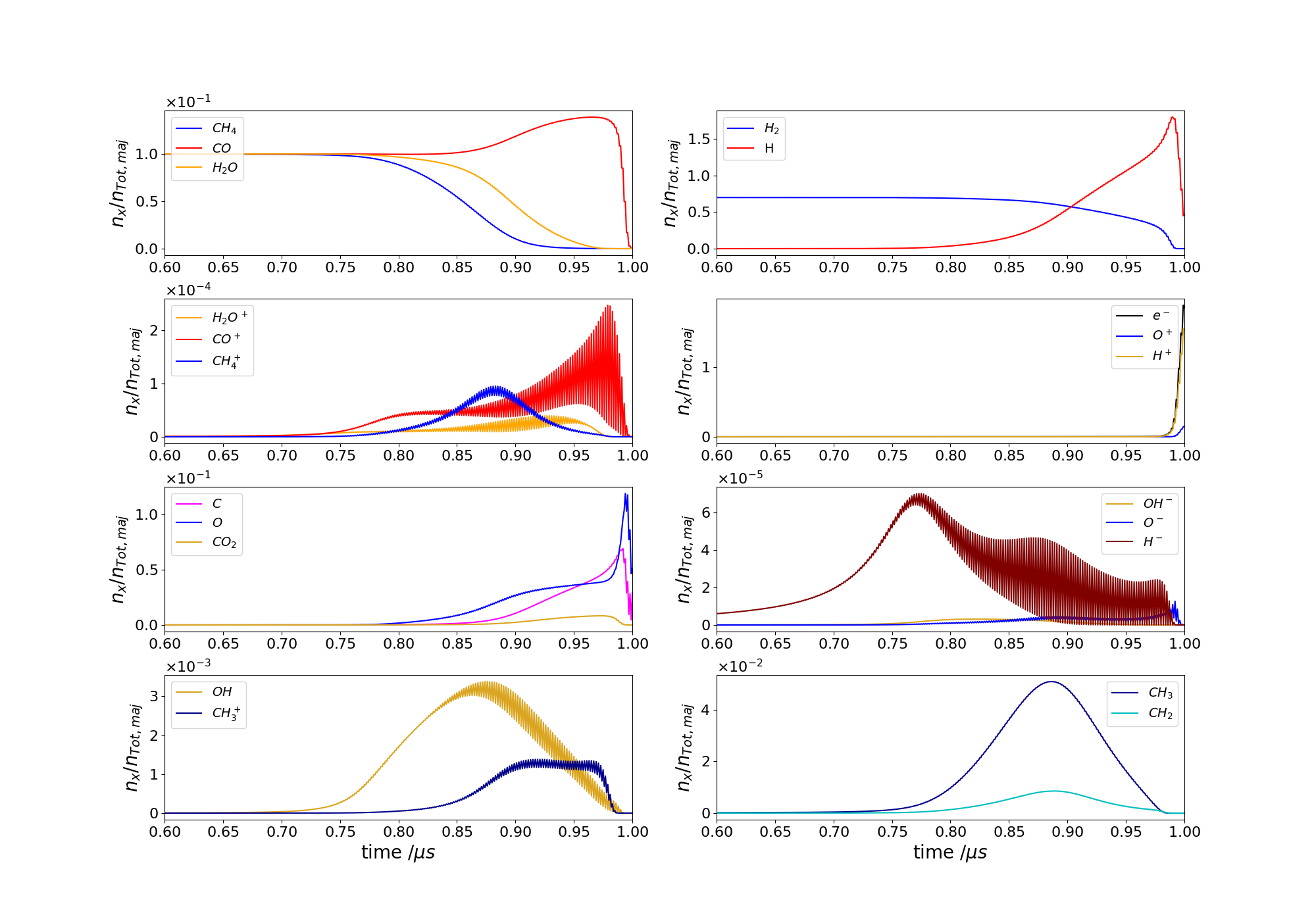}
    \caption{The change in multiple different species in the second multi-pulse scenario (1 ns on-pulse, 1 ns off-pulse) at 3.35eV and a gas temperature of 1600K. Note that early‑time growth is not visible on the linear scale, so the plot begins once the signal becomes resolvable. Displayed on the y axis is the normalised number density of each species, where ${\rm n_x}$ is number density of species x and ${\rm n_{tot,maj}}$ is the combined number densities of ${\rm H_2, CH_4, H_2O}$ and CO. The overall trends of destruction and construction are very similar to the singe-pulse scenario. The staggered addition of electrons and rapidly switching energy can be seen in the evolution of ionised and anion species and at smaller energies. But for only a slightly increased electron energy of 3.35eV, the regular sparking would lead to the same complete exhaustion of initial species seen as in the single-pulse case.}
    \label{fig:mplotpls500}
\end{figure*}
%%%%%%%%%%%%%%%%%%%%%%%%%%%%%%%%%%%%%%%%%%%%%%%%%%%%%%%%%%%

\section{Results \& Discussion}
\label{sec:results}

\begin{figure}
    \centering
    \includegraphics[width=0.99\linewidth]{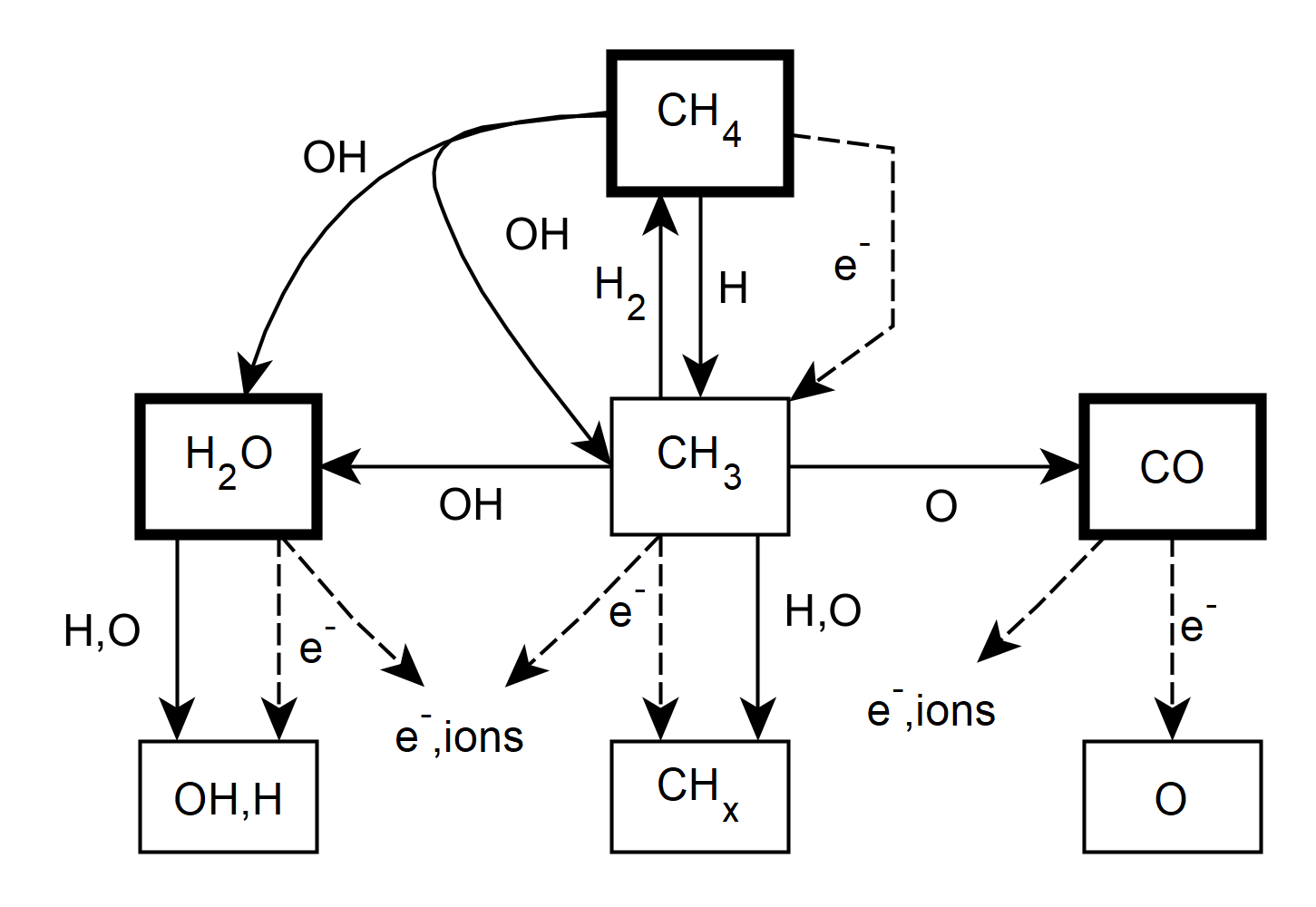}
    \caption{The most significant reaction pathways for our modelling of electrically-active chemistry in a transition dwarf atmospheric composition.
    The pathway from one major molecules to the next is shown by each arrow with the other reactant indicated next to it.
    Neutral-neutral reactions are indicated by bold arrows, electron-neutral reactions by dashed arrows.
    The network highlights the main results of electron dissociation upon our initial compositions of ${\rm CH_4, H_2O}$ and CO (whose boxes are highlighted as the primary species we are interested in) and how the resulting dissociated radicals and atomic species combine to promote growth in ${\rm H_2O}$ and CO.}
    \label{fig:flow}
\end{figure}

\begin{figure*}
    \centering
    \includegraphics[width=0.99\linewidth]{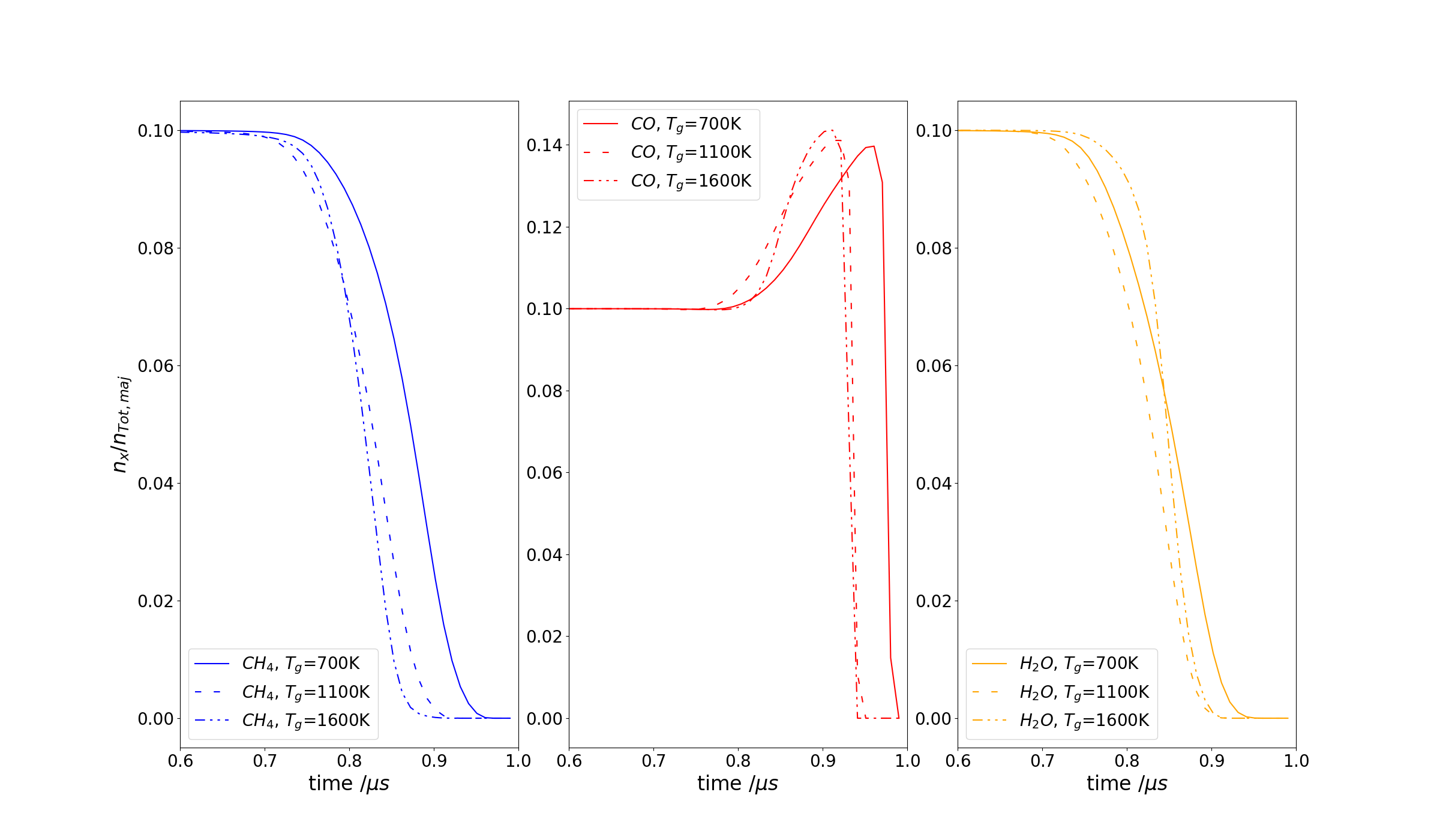}
    \caption{\textit{Left}: The change in normalised number density of key species ${\rm CH_4}$, ${\rm H_2O}$ and CO at an electron temperatures of 3.05eV at three different gas temperatures (700K, 1100K and 1600K).
    Note that early‑time growth is not visible on the linear scale, so the plot begins once the signal becomes resolvable.
    Displayed on the y axis is the normalised number density of each species, where ${\rm n_x}$ is number density of species x and ${\rm n_{tot,maj}}$ is the combined number densities of ${\rm H_2, CH_4, H_2O}$ and CO.
    As gas temperature increases we see neutral-neutral reactions between radical and atomic products of electron-dissociation lead to the reformation of ${\rm H_2O}$.
    Where ${\rm H_2O}$ initially is destroyed at a faster rate than ${\rm CH_4}$, this therefore reverses with increasing gas temperature until it is ${\rm CH_4}$ that is destroyed earlier.
    This indicates a small increase in ${\rm CH_4}$ destruction in hotter layers of an atmosphere due to electrically-active chemistry, potentially a contributor to the shift from ${\rm CH_4}$ lines to CO lines seen at the L-T transition.
    There is also clearly varying destruction rates of each molecule across the three gas temperatures caused by duelling neutral-neutral dissociation and recombination reactions between for these three molecules, fully discussed in their relevant subsections.}
    \label{fig:majplots}
\end{figure*}

Figures \ref{fig:mplotsingtg}, \ref{fig:mplotpls1} and \ref{fig:mplotpls500} present a representative example of the electrically activated chemistry produced by our simulations for all three scenarios, illustrating its influence on the key CO-CH$_{4}$ pathways in a substellar atmosphere using the 0-D substellar plasma activation \& reaction
kinetics code \texttt{SPARCKS}.
It captures the typical modification of the CO-CH$_{4}$ chemistry observed across the parameter ranges $T_{\rm gas} \in [700, 1600]\,\mathrm{K}$ and $T_{e} \in [2, 5]\,\mathrm{eV}$ in contrast to neutral-chemistry in the absence of any electrical activation, with no injected electrons.
In neutral-chemistry, there is no present electron or ion chemistry, showing negligible changes in species abundance over the timescales of our modelled scenarios.
In contrast, when the electron-moderated chemistry is introduced we see widespread change in species abundance as it prompts dissociation, ionization and recombination.
The major reactions of the chemical model is shown in Figure \ref{fig:flow} and are present for both a single discharge and multiple sparking-like events.
In general, the initial atmospheric composition sees electrically-activated reactions dominate, with the growth rate of the population of radicals and atomic gases (such as CH$_{3}$ and atomic H and O) as the result of electron-dissociation. This behaviour correlates with the electron number density.
As electron-impact generated species interact with H$_{2}$O and CO (Figure \ref{fig:flow}), the electrons indirectly control the contrasting rate of destruction of CH$_{4}$ and H$_{2}$O and the growth in CO. 
This leads to a high correlation between all the species on our modelling and the electron number density.
In Figure \ref{fig:mplotsingtg}, we demonstrate where this behaviour dominates as we reach an electron energy of 3.05eV, causing significant change to the abundance of all species.
The increased energy directly leads to more dissociation and ionization, which leads to a growth in CO as the bulk of CH$_{4}$ and H$_{2}$O are dissociated and ionized at 3.05eV.
Reducing/increasing the electron energy results in this behaviour being reduced/increased, taking more/less time to reach the levels seen in Figure \ref{fig:mplotsingtg}.
There is also a variance in number densities due to gas temperature shown for CH$_{4}$, H$_{2}$ and CO in Figure \ref{fig:majplots}, that we will discuss in the relevant sections for each molecule.
The effect of increasing electron energy is non-linear as shown in Figures \ref{fig:ch4co1} - \ref{fig:ch4co3}, with higher‑energy electrons producing disproportionately larger chemical deviations from the neutral baseline.
For both scenarios with multiple sparks we see the same overall behaviour with significant exhaustion of initial species seen at 4.0eV for scenario 1 (seen in Figure \ref{fig:mplotpls1}) and at 3.35eV for scenario 2 (seen in Figure \ref{fig:mplotpls500}).
Without CH$_{4}$ and H$_{2}$O to preferentially dissociate and ionize, more resilient species like CO and H$_{2}$ will be targeted and we see an increase in atomic and then ionized species.
At lower electron temperatures, we begin to see similar behaviour and if the discharge was to last longer than 1 microsecond, we would likely see the same trends of destruction and growth.
In subsequent sections we will go into detail on the trends followed by the most significant species in the simulation.

In the absence of the free electrons, to dissociate CH$_{4}$, H$_{2}$O and CO would only be achieved with either very high gas temperatures or very high initial number densities of dissociating atomic or radical species.
Electron-neutral dissociation and the ionization that increases electron number density are thus the twin drivers to the impact of the electrically-active chemistry.
The rate of this change varies with electron temperature, with a greater given energy to the electrons naturally resulting in a greater impact.
The pulsed scenarios require higher electron energies to obtain comparable results to the single-pulse case due to the shorter on-pulse duration, which is now controlled by the inter-grain interaction time. When sufficient energy is input over an appropriate timescale, the total dissociation and ionization of the atmospheric species occurs as seen in Figures \ref{fig:mplotpls1} and \ref{fig:mplotpls500}.

The simulated gas temperature is a secondary driver of the behaviour of atmospheric species.
Many rate coefficients for neutral-neutral and neutral-ion reactions in our network are dependent on gas temperature.
Although the initial atmospheric make-up is stable to these reactions, electron-neutral dissociation breaks up stable molecules into more reactive species.
Their reactions are then important contributors to the modelled behaviour as seen in Figure \ref{fig:flow}. 
CH$_{3}$, atomic hydrogen (H), atomic oxygen (O) and hydroxyl radicals (OH) in particular go on to serve moderating roles in both the rate of destruction of CH$_{4}$, H$_{2}$O and CO, and the generation of electrons.

Figure \ref{fig:majplots} shows a single pulse run of the simulation at three different gas temperatures.
We see there have been significant reductions in ${\rm CH_4}$ and then ${\rm H_2O}$.
This is accompanied by a small rise in CO.
There is also a clear difference in the onsets of these behaviours between the different gas temperatures, variations which also occur for both multi-pulse scenarios.
In the following subsections we go into further detail on the behaviour of these and other significant species in the simulation.

\subsection{${\rm \bf CH_4}$}
${\rm CH_4}$ undergoes dissociation into smaller hydrocarbons, declining significantly over time as clearly shown in the upper left segment of our figures.
The primary driver of this is dissociation via the electrically-activated reaction ${\rm e^- + \rm CH_4 \rightarrow CH_3 + H + e^-}$.
There are also a significant number of neutral-neutral reactions which contribute as well, causing the gas-temperature dependent variation in number density visible in Figure \ref{fig:majplots}, with the increased destruction rate of C$H_{4}$ and decreased rate of destruction for H$_{2}$O with increased gas temperature.
These primarily include ${\rm CH_4 + H \rightarrow CH_3 + H_2}$ and ${\rm CH_4 + OH \rightarrow  CH_3 + H_2O}$.
This results in ${\rm CH_4}$ exhausting itself at a rate greater than any of the other initial gases in our simulations, even ${\rm H_2O}$ which initially decreases at a quicker rate due its primary electron-dissociation reaction having a higher calculated rate coefficient.
A small amount of ${\rm CH_4}$ also undergoes dissociative ionization to ${\rm CH_3^+}$, however the main source of ${\rm CH_3^+}$ is through the ionization of dissociated ${\rm CH_3}$.
${\rm CH_3}$ as shown in Figure \ref{fig:flow} is a crucial stage in the overall behaviour of our atmospheric mix when exposed to the chemistry of energetic electrons.
The ionization to ${\rm CH_3^+}$ is one of the primary factors in increasing electron number density as will be discussed in the subsection on electrons.
Due to electron-dissociation being the primary pathway for ${\rm CH_4}$, this reliance on ${\rm CH_3}$ ionization causes the number densities of both molecules and electrons to correlate strongly whatever the gas temperature as can be seen in Figures \ref{fig:mplotsingtg}, \ref{fig:mplotpls1} and \ref{fig:mplotpls500}.
Removing both these sources of ${\rm CH_3^+}$ results in a 10-times factor decrease in the electron number density across our range of electron temperature. 
A higher electron temperature would therefore be needed to achieve similar disruption of key molecules within our simulation timeframe.
The electron-dissociation of ${\rm CH_3}$ into smaller hydrocarbons (and eventually C) also disrupts the reformation cycle between ${\rm CH_4}$ and CO.

There is also variation of ${\rm CH_4}$ number density with gas temperature.
This is caused by a few crucial reactions that vary with gas temperature.
${\rm CH_4}$ and its dissociated hydrocarbons react with species such as H, O and OH to either dissociate them or combine to form molecules such as ${\rm H_2O}$ and CO.
However, there are further gas temperature-dependent reactions, such as ${\rm H_2 + OH \rightarrow \rm H_2O + H}$, that further control these reactants.
Therefore at different electron temperatures ${\rm CH_4}$ destruction can be either quicker or slower at higher gas temperatures.

\subsection{${\rm \bf H_2O}$}

${\rm H_2O}$ molecules in the simulation primarily undergo dissociation into either ${\rm H_2}$ and O or OH and H, with declining abundance visible in our figures.
These products react critically with ${\rm CH_4}$ and its products as well as ${\rm H_2}$ to influence their rates of destruction, applying the gas temperature dependence mentioned in the previous section.
${\rm H_2O}$ is also critical in the initial increase in electron number density, through ionization to ${\rm H_2O^+}$ in the reaction ${\rm e^- + H_2O \rightarrow  H_2O^+ + 2e^-}$.
This initial increase (after the initial injection(s) of electrons are mostly consumed through reacting with ${\rm H_2}$) acts to slowly increase the dissociation of ${\rm CH_4}$.
This eventually leads to an exponential increase in electron number density through the ionization of ${\rm CH_3}$ (as discussed in the previous section).
When the ionization of ${\rm H_2O}$ is removed from the simulation, similarly to when the reactions for the ionization of ${\rm CH_3^+}$ were removed, there is a 10-times factor decrease in generated electron number density for a particular electron and gas temperature.
${\rm H_2O}$ itself is a crucial component of the activity we model through its dissociation products and interactions with electrons.
When H$_{2}$O is not included in the initial atmospheric composition, there is far less reactivity with minimal dissociation of ${\rm CH_4}$ over the duration of our simulations and a negligible recovery in electron number density from its initial consumption.

${\rm H_2O}$ itself undergoes electron-dissociation at a faster rate than ${\rm CH_4}$ as can be seen in the 700K case in Figure \ref{fig:majplots}.
However, at higher temperatures this trend reverses with the destruction rate of ${\rm CH_4}$ overtaking ${\rm H_2O}$ above 1100K.
This is due to the gas temperature dependent reformation of ${\rm H_2O}$ from the combination of hydrocarbons and OH as shown in Figure \ref{fig:flow}.
Decreasing the gas temperature decreases the rate of this reaction and, if the electron temperature (or electron number density) isn't high enough to dissociate OH or hydrocarbons too quickly, the higher reaction rate for H$_{2}$O electron dissociation will cause its number density to decrease faster than ${\rm CH_4}$.

\subsection{CO}

Carbon monoxide also dissociates but at a far slower rate than ${\rm CH_4}$ and ${\rm H_2O}$ as shown in our figures by its minimal rate of destruction, a behaviour that persists until the increasing destruction of ${\rm H_2O}$.
This is due to the \texttt{BOLSIG+}-derived rate coefficients for its electron-moderated chemistry being significantly smaller, a results of both a smaller cross-section and its cross section peak occurring at a higher energy.
The primary electron-neutral reaction is electron dissociation due to the reaction ${ \rm e^- + CO \rightarrow C + O + e^-}$.
There is a small amount of ionization through ${\rm e^- + CO \rightarrow CO^+ + 2e^-}$.
${\rm CO^+}$ is also created by ion-neutral reactions such as ${\rm C^+ + OH \rightarrow CO^+ + H}$, but the majority of this ${\rm CO^+}$ experiences recombinative dissociation through ${\rm CO^+ + e^- \rightarrow C + O}$.
Therefore, the main impact of the injected electrons on CO itself is to dissociate to C and O.

The slow destruction of CO can also be attributed to its lack of neutral-dissociation reactions.
CO instead only combines in our simulations, including with the aforementioned products of ${\rm CH_4}$ and ${\rm H_2O}$ dissociation shown in Figure \ref{fig:flow}.
Through reacting with ${\rm O^-, O^*}$ and OH it can create ${\rm CO_2}$, leading to a gradual but steady increase in the number density of the molecule.
Eventually through electron, ion and neutral dissociation this will mostly convert back to CO.
CO is  stable against neutral dissociation with no reactions being found in the literature without the presence of a third reactant.
This, along with the combination of hydrocarbons with O, leads to the CO number density recovering and eventually increasing past its starting abundance with large enough numbers of dissociated ${\rm CH_x}$ and atomic O combine.
This creates the small peaks in CO visible in our figures before an exponentially increasing electron number density and the lack of more reactive species to ionize eventually causes it to dissociate and exhaust similar to ${\rm CH_4}$ and ${\rm H_2O}$.
Overall,our simulation shows CO to survive longer and even benefit off electrically-activated chemistry in contrast to ${\rm H_2O}$ and ${\rm CH_4}$.

\subsection{Electrons}

The initial electrons injected into the simulations mainly perform dissociative attachment via the reaction ${\rm H_2 + e^- \rightarrow H + H^-}$.
This results in the electron number density falling to around a hundredth of its initial population, where a balance is then struck between this reaction and it's reverse, ${\rm H + H^- \rightarrow H_2 + e^-}$, forming a steady small population of ${\rm H^-}$ as seen in Figure \ref{fig:mplotsingtg}.
In the multi-pulse scenarios this is similar with the behaviour persisting as long as the high electron temperature is applied with the number density visibly oscillating depending on the pulse frequency, as seen in Figures \ref{fig:mplotpls1} and \ref{fig:mplotpls500}.
This then increases as the molecules shown in \ref{fig:flow} ionize.
The electron ionization of H$_{2}$O, mainly into ${\rm H_2O^+}$, first slowly increases the number of electrons.
This is slowly at first as the principle electron-moderated reactions are dissociation of the initial atmospheric constituents, into H, O, OH and ${\rm CH_3}$.
The ionization of ${\rm CH_3}$ to ${\rm CH_3^+}$ is quicker than ${\rm CH_4}$ to ${\rm CH_4^+}$ and H$_{2}$O into ${\rm H_2O^+}$.
Therefore, as more ${\rm CH_3}$ is dissociated, we see a greater amount of ionization and the electron number density increases at a more rapid pace.
This growth gradually dominates the behaviour of the simulation.
More electrons cause increased dissociation of ${\rm CH_4}$ to ${\rm CH_3}$ which in turn prompts further ionization in a feedback loop-style process.
The increase in electron number density then also increases the dissociation and ionization of all other species in the simulation.
All ionized species, combined molecules and dissociated products therefore correlate strongly with the electron number density, the only exceptions seen in Figures \ref{fig:mplotsingtg} and \ref{fig:mplotpls500} being the anions which have relatively stable median abundances.
This continues until the population of electrons become so large that they dissociate and ionize every molecule into their atomic components, e.g. what would be expected in the superheated centre of a lightning discharge.

For the pulsed cases as can be observed in Figures \ref{fig:mplotpls1} and \ref{fig:mplotpls500}, the slower injection of electrons along with periods of on-pulse and off-pulse leads to periods of growth through ionization and recombination.
Within the timescales of each pulse there is not enough recombination to offset the growth. 
This leads to the same growth of electron number density as in the single-pulse case, in-turn dominating the simulation through its dissociated and ionized products.
As long as the rate of ionization outpaces the recombination off-pulse, we see the behaviour that would cause the persistent sparking behaviour we are interested in.
In this case, even smaller electron energies than displayed would be given the time necessary to create significant changes to the chemical balance of the atmosphere.

Some of the dissociated products as mentioned above contribute to dissociation and ionization themselves.
This noticeably leads to a slight variation in electron number density with gas temperature, which would otherwise be unexpected to effect the behaviour of non-thermal electrons.
With electron number density tied to the amount of ionization present in the atmosphere versus attachment, the key role of ${\rm CH_3}$ ionization (whose production as mentioned above and shown in Figure \ref{fig:flow} is boosted by neutral-neutral dissociation) boosts this ionization, electron number density and therefore the speed at which dissociation processes occur.
At lower electron temperatures this leads to between 1000 K and 1100 K having the highest amount of electrically-activated chemistry by the end of the simulation, whereas at higher electron temperatures the increased ionization of ${\rm CH_3}$ eventually allows for increased gas temperature-dependent neutral-neutral dissociation.

The lack of certain ionization reactions or H$_{2}$O itself can decrease the speed of the dissociation and ionization caused by the electrically-active chemistry.
${\rm H_2O}$ and ${\rm CH_3}$ ionization are crucial as is the presence of the dissociation products of H$_{2}$O.
Without any initial injected electrons there is zero measurable change in the atmospheric composition of our simulations over our reaction timescales.
The initial dissociation of our starting atmospheric species is what helps to drive other dissociation reactions.
To see the same impact of electron dissociation from a neutral-neutral source such as H dissociation for the single-pulse case requires 10,000 times more H than the initial number density of injected electronsn ($10^{-6}$ for our single-pulse case).
Without a doubt, the population of electrons and the highly reactive plasma chemistry they engender are capable of dramatically changing atmospheric make-up over shorter timescales than is possible for other mechanisms using physically plausible atmospheric make-ups.

\subsection{${\rm \bf H_2}$}

${\rm H_2}$ is set at 70 \% of our atmosphere and would be the majority component of an L/T transition brown dwarf atmosphere.
The main interests of our simulations are in electrically-active chemistry with the other main atmospheric components being crucial to the L-T transition.
But, ${\rm H_2}$ as the most common species in our model atmosphere will also interact the most with the injected electrons. 
${\rm H_2}$ undergoes dissociative attachment through ${\rm H_2 + e^- \rightarrow H^- + H}$, but this is mostly balanced from its reverse reaction, with only a small amount of the dissociated ${\rm H^-}$ and H eventually becoming ${\rm H^+}$.
Electrons react more strongly with ${\rm H_2O}$, and ${\rm CH_x}$ molecules due to smaller reaction cross sections with higher energy peaks.
${\rm H_2}$'s strongest impact is instead through neutral-neutral reactions with OH and O, through which it impacts the population of ${\rm H_2O}$ and the gas temperature dependence we see in the simulation from dissociated H.
It is only with increased electron number density that dissociation processes or attachment of ${\rm H^-}$ begin to make significant dents in the number density of ${\rm H_2}$.
Like CO, it is one of our less reactive molecules and with high enough electron populations would exhaust at a slower rate than ${\rm CH_4}$ and ${\rm H_2O}$.
So to see significant changes, higher electron temperatures are required.

\subsection{CH$_{4}-$CO ratio}

\begin{figure}
    \centering
    \includegraphics[width=0.99\linewidth]{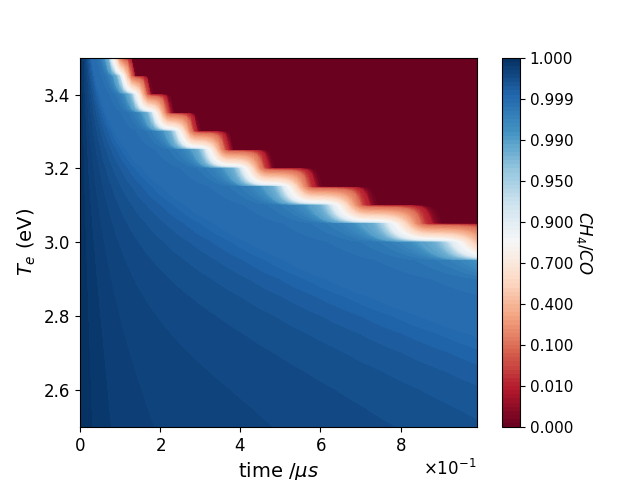}
    \caption{The change in the ratio of the number densities of ${\rm CH_4}$ to CO over time at different electron temperatures for a discharge of one microsecond and a gas temperature of 1600K. The electrically-active chemistry at even small energies are able to affect large changes in this ratio, with the ratio fully halved at 3.0eV after one microsecond.}
    \label{fig:ch4co1}
\end{figure}

\begin{figure}
    \centering
    \includegraphics[width=0.99\linewidth]{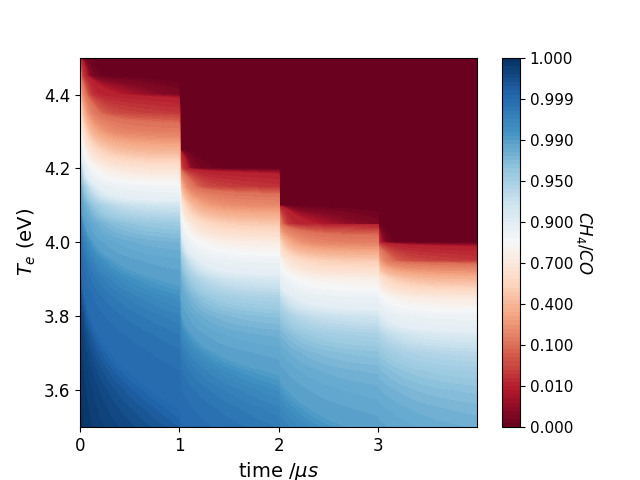}
    \caption{The change in the ratio of the number densities of ${\rm CH_4}$ to CO over time at different electron temperatures and a gas temperature of 1600K for a pulsed discharge of 10 nanoseconds every one microsecond. These short pulses and the electrically-active chemistry that they promote are at high energies capable of promoting CO to the expense of ${\rm CH_4}$ in the same way as one continuous pulse.}
    \label{fig:ch4co2}
\end{figure}

\begin{figure}
    \centering
    \includegraphics[width=0.99\linewidth]{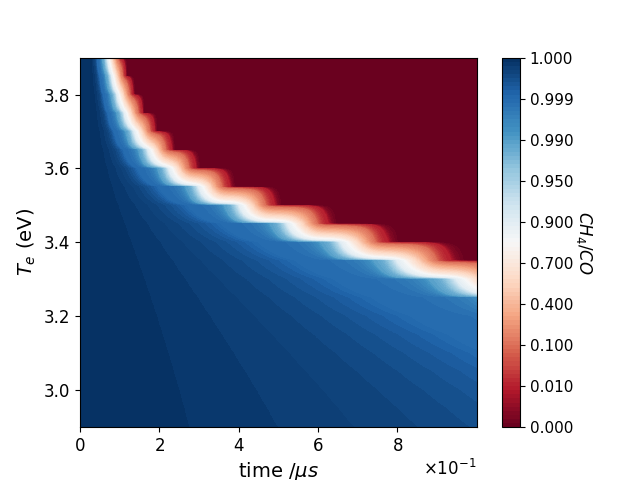}
    \caption{The change in the ratio of the number densities of ${\rm CH_4}$ to CO over time at different electron temperatures and a gas temperature of 1600K for a pulsed discharge of 1 nanosecond for every two nanoseconds. The persistent discharges is enough to recreate the behaviour of a discharge for one full microsecond, for a slightly higher energy.}
    \label{fig:ch4co3}
\end{figure}

Our results primarily show the promotion of CO over ${\rm CH_4}$ in the presence of ${\rm H_2O}$, with the ${\rm CH_4}$/CO ratio halving in one microsecond with only 3.0eV applied.
This aligns with the changing strength of these molecules over the L-T transition.
This behaviour can be seen occurring in all three tested cases, with the single-pulse scenario in Figure \ref{fig:ch4co1}, requiring less energy to do so than the pulsed scenarios in Figures \ref{fig:ch4co2} and \ref{fig:ch4co3}.
The step-like nature of the single-pulse case is due to significant variation between energies in our step-size of 0.05 eV and would be expected to smooth out given a finer energy range.
When coupled with the proposed changes in cloud coverage from cloudy late L-dwarfs to cloudless early T-dwarfs, this shows that cloud-dependent electrically-activated chemistry could provide a crucial role in behaviour at the transition.
As a disruptor in conventional chemistry and a potential genesis for life's initial building blocks, accounting for these processes will be key for non-LTE atmospheric modelling and observation.
Lightning and its ubiquitousness in the galaxy across the substellar regime is a question that must be resolved as we continue the hunt for life into ever greater precision.

\section{Conclusions}
\label{sec:discussion}

In this paper, we have developed the first model capturing the effects of electrically-activated chemistry upon key chemistry in a transition brown dwarf atmosphere, showing its effect on key CO-CH$_{4}$ chemistry. We present how it is physically consistent with the behaviour of the CO/CH$_{4}$ transition, using our \texttt{SPARCKS} simulations to demonstrate how electrically activated chemistry can influence key molecular abundances at the L–T transition in brown dwarfs. In L-dwarfs, abundant clouds sustain charging and discharging, driving plasma chemistry that preferentially destroys CH$_{4}$ and produces CO. Transitioning from L to T dwarfs, where clouds dissipate, electrical activation is quenched and the plasma pathways are disrupted. Without this activated chemistry, CH$_{4}$ can re-emerge as the dominant species due to conventional chemistry while CO declines. 

The main point demonstrated by our results is that given a discharge of enough energy, a relatively low number of electrons can have a significant impact on local atmospheric chemistry, in our case halving the  CH$_{4}$ to CO ratio with only a microsecond long discharge at 3eV.
Electrically-active plasma chemistry is able to significantly alter atmospheric chemistry through ionization and dissociation reactions at small timescales, due to the differing reactivity of major species in an atmosphere.
Ionization reactions overpower recombination due to reactive ionizations of H$_{2}$O and dissociated radicals such as ${\rm CH_3}$.
All of this results in a significant chemical impact that would be impossible without the electron-moderated pathways made possible by plasma chemistry.

The total impact of this chemistry upon a substellar atmosphere, requiring an electric discharge thus would be limited by factors such as lightning frequency and cloud coverage.
However, other than large scale discharges resulting from the combination of triboelectric charging and charge separation in extensive dust cloud structures, there can also be smaller-scale discharges.
 Small scale discharges, such as inter grain microdischarges, occur when triboelectric charging leads to charged grains coming into close proximity.
 The frequency of these events depends on environmental factors including the dust particle size distribution function, the dust particle number density, and the prevailing dust dynamics.
Represented by the pulsed experimental cases, this can result in a continual production of electrons, ions and radicals as long as enough energy remains in the atmosphere to avoid the domination of recombination processes over ionization. In general, electrical discharges generate ions, radicals, excited states, and charged dust (beyond what is included in our present model) that do not vanish immediately after the event. At the pressures and temperatures typical of brown dwarf cloud layers, recombination and neutralisation can be inefficient, as metastable and vibrationally excited species have long lifetimes, and radical chemistry proceeds slowly. Dust grains further store charge and reactive species and release them over extended periods, while vertical mixing is slower than many chemical pathways. Consequently, the post discharge gas relaxes only gradually, and the discharge generated species persist well beyond the discharge itself rather than instantly returning to equilibrium. Quantifying this behaviour requires a spatially dependent model that includes dust grains and transport processes; this is currently being developed by the authors.

The plasma activation and reaction kinetics model currently focuses on key species, but incorporating more complex hydrocarbons and amino acids would open the door to exploring the broader impact of electrical activation.
One possible significant product not included was the ${\rm C_2H_x}$ hydrocarbons.
As a significant product of reactions with dissociated ${\rm CH_3}$, there would likely be a generated population of ${\rm C_2H_x}$ hydrocarbons within the simulation.
However, as we determined that this would for the most part lead to eventual electron-dissociation into ${\rm H_2}$ and then H, we felt we could exclude these molecules for the sake of simulation speed given that the reaction leads to the same result as dissociating into smaller hydrocarbons.
With further research we could include these molecules and other complex molecules such as amino acids, both to get an idea of the overall contribution towards different chemical pathways from the ${\rm CH_3}$ radical, and to explore the amounts of potentially life-creating complex chemistry is created by the electrically-active chemistry.

In our modelled atmospheres we have sought to represent a brown dwarf at the L/T transition.
From our results electrons, ions and radicals are produced in the path of the discharge before processes like diffusion, convection, settling and adsorption would then carry these components into the wider atmosphere.
The chemical impact on the whole atmosphere would then depend on the frequency and survival of dust-to-dust microdischarges and the total cloud coverage in the brown dwarf.
Given patchier clouds from L- to T- dwarfs, the impact of electrically-active chemistry on a brown dwarf would decrease as we transition to cloud-free atmospheres.
As our results show a greater impact from electrically-active chemistry to ${\rm CH_4}$ than CO, we can infer that there would be a lower ${\rm CH_4}$-to-CO ratio for a transition dwarf with this chemistry than one without.
With the L-T transition also characterised by the shift from strong CO lines to strong ${\rm CH_4}$ lines from L-dwarfs to T-dwarfs, this chemistry could be very significant in how brown dwarf populations evolve across time as they cool.
Coupling this chemistry to a more detailed physical model whether 1-D or a full General Circulation Model (GCM) would a highly desirable future research direction.
This would be especially useful given how particle and cloud dynamics considered partially responsible for the transition \citep{2019Tremblin,2025Teinturier} would also heavily affect grain dynamics and transport.

One way to observe any modelled effect would be to compare differences in transition dwarf atmospheres with different abundances of molecules key to electrically-active chemistry.
H$_{2}$O, a primary component of brown dwarf atmospheres at the L-T transition, is such a key species to the results  of our simulation, without which the observed dissociation and ionization of ${\rm CH_4}$ takes place at a much slower pace.
H$_{2}$O holds a potential for observation that other important key molecules such as the short-lived ${\rm CH_3}$ radical do not.
The observation of similar brown dwarfs with varying observed abundances of H$_{2}$O in their atmospheres could indicate cases where otherwise similar objects could have differing amounts of electrically-activated chemistry.
If this lines up with spectral characterisations of ${\rm CH_4}$ and CO, and observations on particle size, then this could be another way of observing the impact electrically-activated chemistry at the L-T transition.

If plasma chemistry is as key to the L-T transition as our modelling suggests it could be, it naturally leads to how essential it may be to other transition spaces in substellar astrophysics.
In future efforts we hope to include Nitrogen bearing species to our code, allowing us to reach down to the parameter space of the T-Y transition where ${\rm NH_3}$ is a key molecule.
Given the molecule HCN has been proposed as a marker of lightning activity and started to be observed via JWST observations \citep{2025Matthews}, this could also allow the prediction of the strength of lightning-activity markers, leading to a first step in observing lightning outside our solar system.
Continuing down to the terrestrial, where electrically-activated chemistry has been proposed to be a prompter of prebiotic chemistry, the incorporation of our chemistry into atmospheric models could allow a tentative suggestion of the likelihood for an atmosphere to generate amino acids considered key to the formation of primordial life.

\section{Acknowledgements}
CRS and MIS are grateful for funding from STFC via Grant Number ST/X000885/2. This work makes use of the SPARCKS code, developed and maintained by the Astronomy \& Astrophysics at the University of Glasgow. Access and further development are coordinated through the originating group.
We thank the the anonymous referee for their constructive and timely comments which have improved the paper in many aspects.

%%%%%%%%%%%%%%%%%%%%%%%%%%%%%%%%%%%%%%%%%%%%%%%%%%
\section{Data Availability}

The data underlying this article is available for all three scenarios at http://dx.doi.org/10.5525/gla.researchdata.2226.
The cross sections and rate coefficients used in our simulations can be found in the Appendix.

%%%%%%%%%%%%%%%%%%%% REFERENCES %%%%%%%%%%%%%%%%%%

% The best way to enter references is to use BibTeX:

\bibliographystyle{mnras}
\bibliography{example} % if your bibtex file is called example.bib

% Alternatively you could enter them by hand, like this:
% This method is tedious and prone to error if you have lots of references
%\begin{thebibliography}{99}
%\bibitem[\protect\citeauthoryear{Author}{2012}]{Author2012}
%Author A.~N., 2013, Journal of Improbable Astronomy, 1, 1
%\bibitem[\protect\citeauthoryear{Others}{2013}]{Others2013}
%Others S., 2012, Journal of Interesting Stuff, 17, 198
%\end{thebibliography}

%%%%%%%%%%%%%%%%%%%%%%%%%%%%%%%%%%%%%%%%%%%%%%%%%%

%%%%%%%%%%%%%%%%% APPENDICES %%%%%%%%%%%%%%%%%%%%%

\appendix

\section{Rate Coefficients}
\label{sec:appendix}

\begin{table*}
    \caption{The CHO reactions included in SPARCKS are displayed here along with their rate coefficients and source. The rate coefficients are in $cm^3/s$.}
    \label{rcoefftable1}
    $
         \begin{array}{lcrc}
            \hline\hline
            \noalign{\smallskip}
            \text{RNO} & \multicolumn{1}{c}{\text{Reaction}} & \multicolumn{1}{l}{\text{Rate Coefficient} [{\rm cm^3/s}]} & \multicolumn{1}{l}{\text{Source}} \\
            \noalign{\smallskip}
            \hline
            \noalign{\smallskip}
            1 & {\rm e^- + O_2 \rightarrow O_2^+ + 2e^-}      &  {\rm f(\sigma)} & [1]\\ 
            2 & {\rm e^- + O_2 \rightarrow O^- + O}        &  {\rm f(\sigma)} & [1] \\ 
            3 & {\rm O^- + O_2 \rightarrow O_3 + e^-}       &  5\times10^{-15} & [2] \\ 
            4 & {\rm O(3P) + O^- \rightarrow O_2(Sig) + e^-}  &  1.6\times10^{-10} &[2]  \\ 
            5 & {\rm O_2^+ + O_2^- \rightarrow 2O_2}        &  2.01\times10^{-7} (300/T_g)^{1/2}&[2] \\ 
            6 & {\rm e^- + O_3 \rightarrow O^- + O_2}       &  2.12\times10^{-9}T_e^{-1.058}\exp{(-0.93/T_e)}  &[2] \\ 
            7 & {\rm e^- + O_3 \rightarrow O_2^- + O}       &  9.758\times10^{-8}T_e^{-1.309}\exp{(-1.007/T_e)} &[2]\\ 
            8 & {\rm e^- + O_3 \rightarrow O_2 + O(3P)+e^-}  &  1\times10^{-8} &[2] \\ 
            9 & {\rm e^- + O_2 \rightarrow O_2^* + e^-}      &  1.37\times10^{-9}\exp{(-2.14/T_e)}&[2] \\ 
            10 & {\rm O^+ + O_2 \rightarrow O + O_2^+}    &  2.0\times10^{-11}(300/T_g)^{1/2}  &[2] \\ 
            11 & {\rm e^- + O_2 \rightarrow 2O(3P) + e^-}  &  {\rm f(\sigma)}  & [1] \\ 
            12 & {\rm O^- + O_2^+ \rightarrow 3O} & 2.6\times10^{-8}(300/T_g)^{0.44}&[2]  \\ 
            13 & {\rm O^- + O_2^+ \rightarrow O(3P)+ O_2} &  2.6\times10^{-8}(300/T_g)^{0.44}&[2]  \\ 
            14 & {\rm e^- + O_2^+ \rightarrow O(3P) + O^*} &  2.2\times10^{-8}/T_e^{0.5} & [2] \\ 
            15 & {\rm e^- + O^- \rightarrow O + 2e^-}      &  5.47\times10^{-8}T_e^{0.234}\exp{(-2.98/T_e)} & [2] \\ 
            16 & {\rm O^- + O_2^* \rightarrow O_3 + e^-}     &  3.3\times10^{-11} &[2] \\ 
            17 & {\rm O^- + O_2^*\rightarrow O_2^- + O(3P)} &  1.1\times10^{-10} & [2]  \\ 
            18 & {\rm e^- + O_2 \rightarrow O(3P)+ O^+ +2e^-} &  {\rm f(\sigma)} & [1]      \\ 
            19 & {\rm e^- + O_2 \rightarrow O^- + O^+  + e^-}   & 7.1\times10^{-11}T_e^{0.5}\exp{(-17/T_e)} &[2]      \\ 
            20 & {\rm e^- + O_2^*\rightarrow e^- + O_2}      & 2.06\times10^{-9} \exp{(-1.163/T_e)} &[2]        \\ 
            21 & {\rm O_3 + O \rightarrow 2xO_2}        & 1.8\times10^{-11}\exp{(-2300/T_g)}   &[2]          \\ 
            22 & {\rm e^- + O_2^*\rightarrow O + O^-}     & 4.19\times10^{-9}T_e^{-1.376}\exp{(-5.19/T_e)} &[2]  \\ 
            23 & {\rm O_2^*+ O \rightarrow O_2 + O}     & 4.0\times10^{-16}  &[2]                           \\ 
            24 & {\rm O_3 + O_2 \rightarrow 2O_2 + O}    & 7.3\times10^{-10}\exp{(-11400/T_g)}  &[2]         \\ 
            25 & {\rm e^- + C^+ \rightarrow C}          &  4.73\times10^{-12}T_e^{-0.61} & [3]    \\ 
            26 & {\rm O_2^*+ O_3 \rightarrow O + 2O_2}    &  6.01\times10^{-11} \exp{(-2853/T_g)} &[2] \\
            27 & {\rm O^+ O_2 \rightarrow O + O_2}      &  2.56\times10^{-11} \exp{(67/T_g)} &[2] \\ 
            28 & {\rm e^- + O^*\rightarrow e^- + O}       &  8.17\times10^{-9} \exp{(-0.4/T_e)} &[2] \\ 
            29 & {\rm O^*+ O \rightarrow 2O}          &  8\times10^{-12} &[2] \\ 
            30 & {\rm O_2^- + O_2^*\rightarrow 2O_2 + e^-}  & 2\times10^{-10} &[2] \\ 
            31 & {\rm O_2^- + O \rightarrow O^- + O_2}     &  3.31\times10^{-10} &[2] \\ 
            32 & {\rm O_2^- + O^+ \rightarrow O_2 +O(3P)} & 2.7\times10^{-7}(300/T_g)^{1/2} &[2] \\ 
            33 & {\rm O_2^- + O_2^+ >O_2 +2O(3P)} & 1.01\times10^{-7}(300/T_g)^{1/2} &[2] \\ 
            34 & {\rm O^- + O^+ \rightarrow2O(3P)}      & 4.\times10^{-8}(300/T_g)^{0.43}  &[2] \\ 
            35 & {\rm O_2^+ + O_2^-  \rightarrow 2O_2} & 2.01\times10^{-7}(300/T_g)^{0.5}  &[2] \\ 
            36 & {\rm CO_2 + e^- \rightarrow CO_2^+ + 2e^-}       & {\rm f(\sigma)} & [4] \\ 
            37 & {\rm CO + e^- \rightarrow CO^+ + 2e^-}         & {\rm f(\sigma)} & [5]\\ 
            38 & {\rm CO_2 + e^- \rightarrow CO + O^-}          & {\rm f(\sigma)} & [4]\\ 
            39 & {\rm CO + e^- \rightarrow C + O^-}            & {\rm f(\sigma)} & [5]\\ 
            40 & {\rm CO + O^- \rightarrow CO_2 + e^-}          & 5.8\times10^{-9}(T_g)^{-0.4} & [6] \\ 
            41 & {\rm CO_2 + e^- \rightarrow CO + O + e^-}      & {\rm f(\sigma)} & [4]\\ 
            42 & {\rm CO_2^+ + e^- \rightarrow CO + O}          & 4.2\times10^{-7}(T_e/300)^{-0.75} & [6] \\ 
            43 & {\rm CO_2^+ + O \rightarrow O_2^+ + CO}         & 2.6\times10^{-10} & [6] \\ 
            44 & {\rm CO_2^+ + O_2 \rightarrow O_2^+ + CO_2}       & 1.1\times10^{-14}T_g\exp{(904.5/T_g)} & [6] \\ 
            45 & {\rm CO^+ + CO_2 \rightarrow CO_2^+ + CO}       & 1.2\times10^{-9} & [6] \\ 
            46 & {\rm CO^+ + O_2 \rightarrow CO + O_2^+}         & 2\times10^{-10}  & [6] \\ 
            47 & {\rm O^+ + CO_2 \rightarrow O_2^+ + CO}         & 9.4\times10^{-10} & [6] \\ 
            48 & {\rm O + CO \rightarrow CO_2}                & 2\times10^{-20} & [6] \\ 
            49 & {\rm CO_2 + e^- \rightarrow CO^+ + O + 2e^-}    & {\rm f(\sigma)} & [4]\\ 
            50 & {\rm CO_2 + e^- \rightarrow CO + O^+ + 2e^-}    & {\rm f(\sigma)} & [4]\\ 
            51 & {\rm CO + O^- \rightarrow CO_2 + e^-}          & 5\times10^{-10} & [7]  \\ 
            52 & {\rm CO_2 + C \rightarrow 2CO}               & 1\times10^{-15} & [7]  \\ 
            53 & {\rm CO^+ + e^- \rightarrow C + O}            & 6.8\times10^{-7}T_e^{-0.4} &[7]\\ 
            54 & {\rm CH_4 + e^- \rightarrow CH_4^+ + 2e^-} & {\rm f(\sigma)} & [8][9]\\ 
            55 & {\rm CH_4 + e^- \rightarrow CH_3^+ + H + 2e^-} & {\rm f(\sigma)} & [8][9]\\ 
            56 & {\rm CH_4 + e^- \rightarrow CH_2^+ + H_2 + 2e^-} & {\rm f(\sigma)} & [8][9]\\ 
            57 & {\rm CH_4 + e^- \rightarrow CH_3 + H + e^-} & {\rm f(\sigma)} & [8][9]\\ 
            58 & {\rm CH_4 + e^- \rightarrow CH_2 + H_2 + e^-} & {\rm f(\sigma)} & [8][9]\\ 
            59 & {\rm CH_4 + e^- \rightarrow CH + H_2 + H +e^-} & {\rm f(\sigma)} & [8][9]\\ 
            60 & {\rm CH_4 + e^- \rightarrow C + 2H_2 + e^-} & {\rm f(\sigma)} & [8][9]\\ 
            61 & {\rm CH_4 + CH_2 \rightarrow CH_3 + CH_3}  & 3.01\times10^{-19} & [10] \\ 
            \noalign{\smallskip}
            \hline\hline
         \end{array}
    $
\end{table*}

\begin{table*}
    %\caption{The CHO reactions included in SPARCKS are displayed here along with their rate coefficients and source. The rate coefficients are in $cm^3/s$.}
    \label{rcoefftable2}
    $ 
         \begin{array}{lcrc}
            \hline\hline
            \noalign{\smallskip}
            \text{RNO} & \multicolumn{1}{c}{\text{Reaction}} & \multicolumn{1}{l}{\text{Rate Coefficient} [{\rm cm^3/s}]} & \multicolumn{1}{l}{\text{Source}} \\
            \noalign{\smallskip}
            \hline
            \noalign{\smallskip}
            62 & {\rm CH_4 + H \rightarrow CH_3 + H_2}  & 2.18\times10^{-20}T_g^{3}\exp{(-4045/T_g)} & [11] \\ 
            63 & {\rm CH_3 + H_2 \rightarrow CH_4 + H}  & 6.86\times10^{-14}(T_g/298)^2.74\exp{(-39.41/(RT_g))} & [10] \\ 
            64 & {\rm CH_5^+ + CH_2 \rightarrow CH_3^+ + CH_4}  & 9.6\times10^{-10} &[10]  \\ 
            65 & {\rm CH_5^+ + CH \rightarrow CH_2^+ + CH_4}  & 6.9\times10^{-10} &[10]  \\ 
            66 & {\rm CH_5^+ + C \rightarrow CH^+ + CH_4}  & 1.2\times10^{-9} &[10]   \\ 
            67 & {\rm CH_4^+ + CH_4 \rightarrow CH_5^+ + CH_3}  & 1.5\times10^{-9} &[10]  \\ 
            68 & {\rm CH_3^+ + CH_4 \rightarrow CH_4^+ + CH_3}  & 1.36\times10^{-10} & [10]  \\ 
            69 & {\rm CH_2^+ + CH_4 \rightarrow CH_3^+ + CH_3}  & 1.38\times10^{-10} &[10]  \\ 
            70 & {\rm H_2^+ + CH_4 \rightarrow CH_5^+ + H}  & 1.14\times10^{-10} & [10] \\ 
            71 & {\rm H_2^+ + CH_4 \rightarrow CH_4^+ + H_2}  & 1.4\times10^{-9} & [10]  \\ 
            72 & {\rm H_2^+ + CH_4 \rightarrow CH_3^+ + H_2 + H}  & 2.3\times10^{-9}  &[10]  \\ 
            72 & {\rm H^+ + CH_4 \rightarrow CH_4^+ + H}  & 1.5\times10^{-9} & [10]  \\ 
            74 & {\rm H^+ + CH_4 \rightarrow CH_3^+ + H_2}  & 2.3\times10^{-9} &[10]  \\ 
            75 & {\rm H^- + CH_3 \rightarrow CH_4 + e^-}  & 1.0\times10^{-9}& [10]  \\ 
            76 & {\rm e^- + CH_3 \rightarrow CH_3^+ + 2e^-}  & {\rm f(\sigma)} & [12]\\ 
            77 & {\rm e^- + CH_3 \rightarrow CH_2^+ + H + 2e^-}  & {\rm f(\sigma)} & [12]\\ 
            78 & {\rm e^- + CH_3 \rightarrow CH^+ + H_2 + 2e^-}  & {\rm f(\sigma)} & [12]\\ 
            79 & {\rm e^- + CH_3 \rightarrow CH_2 + H + e^-}  & {\rm f(\sigma)} & [12]\\ 
            80 & {\rm e^- + CH_3 \rightarrow CH + H_2 + e^-}  & {\rm f(\sigma)} & [12]\\ 
            81 & {\rm e^- + CH_2 \rightarrow CH_2^+ + 2e^-}  & {\rm f(\sigma)} & [12]\\ 
            82 & {\rm e^- + CH_2 \rightarrow CH^+ + H + 2e^-}  & {\rm f(\sigma)} & [12]\\ 
            83 & {\rm e^- + CH_2 \rightarrow C^+ + H_2 + 2e^-}  & {\rm f(\sigma)} & [12]\\ 
            84 & {\rm e^- + CH_2 \rightarrow CH + H + e^-}  & {\rm f(\sigma)} & [12]\\ 
            85 & {\rm e^- + CH \rightarrow CH^+ + 2e^-}  & {\rm f(\sigma)} & [12]\\ 
            86 & {\rm e^- + CH \rightarrow C^+ + H + 2e^-}  & {\rm f(\sigma)} & [12]\\ 
            87 & {\rm e^- + CH \rightarrow C + H + e^-}  & {\rm f(\sigma)} & [12]\\ 
            88 & {\rm e^- + C \rightarrow C^+ + 2e^-}  & {\rm f(\sigma)} & [13]\\ 
            89 & {\rm e^- + CH_5^+ \rightarrow CH_3 + 2H}  & 2.57\times10^{-7}(T_e/300)^{-0.3}& [10]\\ 
            90 & {\rm e^- + CH_5^+ \rightarrow CH_2 + H_2 + H}  & 6.61\times10^{-8}(T_e/300)^{-0.3}& [10]\\ 
            91 & {\rm e^- + CH_4^+ \rightarrow CH_3 + H} & 1.18\times10^{-8}(T_e/300)^{-0.5}& [10]\\ 
            92 & {\rm e^- + CH_4^+ \rightarrow CH_2 + 2H}  & 2.42\times10^{-8}(T_e/300)^{-0.5}& [10]\\ 
            93 & {\rm e^- + CH_4^+ \rightarrow CH + H_2 + H}  & 1.41\times10^{-8}(T_e/300)^{-0.5}& [10]\\ 
            94 & {\rm e^- + CH_3^+ \rightarrow CH_2 + H}  & 2.25\times10^{-8}(T_e/300)^{-0.5}& [10]\\ 
            95 & {\rm e^- + CH_3^+ \rightarrow CH + H_2}  & 7.88\times10^{-9}(T_e/300)^{-0.5}& [10]\\ 
            96 & {\rm e^- + CH_3^+ \rightarrow CH + 2H}  & 9.0\times10^{-9}(T_e/300)^{-0.5}& [10]\\ 
            97 & {\rm e^- + CH_3^+ \rightarrow C + H_2 + H}  & 1.69\times10^{-8}(T_e/300)^{-0.5}& [10]\\ 
            98 & {\rm e^- + CH_2^+ \rightarrow CH + H}  & 1.0\times10^{-8}(T_e/300)^{-0.5}& [10]\\ 
            99 & {\rm e^- + CH_2^+ \rightarrow C + H_2}  & 4.82\times10^{-9}(T_e/300)^{-0.5}& [10]\\ 
            100 & {\rm e^- + CH_2^+ \rightarrow C + 2H}  & 2.53\times10^{-8}(T_e/300)^{-0.5}& [10]\\ 
            101 & {\rm e^- + CH^+ \rightarrow C + H}   & 3.23\times10^{-8}(T_e/300)^{-0.42}& [10] \\ 
            102 & {\rm CH_3 + H \rightarrow CH_2 + H_2}   & 1.0\times10^{-10}\exp{(-7600/T_g)} & [11] \\ 
            103 & {\rm CH_2 + H \rightarrow CH + H_2}   & 1.0\times10^{-11}\exp{(900/T_g)} & [11] \\ 
            104 & {\rm CH + H_2 \rightarrow CH_2 + H}   & 1.48\times10^{-11}(T_g/298)^{1.79}\exp{(-6.98/RT_g)} & [10]  \\ 
            105 & {\rm CH + H \rightarrow C + H_2}   & 1.31\times10^{-10}\exp{(-6.7/RT_g)} & [10] \\ 
            106 & {\rm C + H_2 \rightarrow CH + H}   & 6.64\times10^{-10}\exp{(-11700/T_g)} & [14] \\ 
            107 & {\rm CH_4^+ + H_2 \rightarrow CH_5^+ + H}   & 4.893\times10^{-11}(T_g/300.d0)^{-0.14}\exp{(36.1d0/T_g)}  & [14] \\ 
            108 & {\rm CH_4^+ + H \rightarrow CH_3^+ + H_2}   & 1.0\times10^{-11} & [10] \\ 
            109 & {\rm CH_2^+ + H_2 \rightarrow CH_3^+ + H}   & 1.6\times10^{-9} & [10] \\ 
            110 & {\rm CH^+ + H_2 \rightarrow CH_2^+ + H}   & 1.2\times10^{-9} & [10] \\ 
            111 & {\rm CH^+ + H \rightarrow C^+ + H_2}   & 7.5\times10^{-10}T_e^{-0.5} & [15] \\ 
            112 & {\rm C^+ + CH_2 \rightarrow CH_2^+ + C}   & 5.2\times10^{-10} & [10] \\ 
            113 & {\rm C^+ + CH \rightarrow CH^+ + C}   & 3.8\times10^{-10} & [10] \\ 
            114 & {\rm e^- + H_2 \rightarrow H_2^+ + 2e^-}  & {\rm f(\sigma)} & [16]\\ 
            115 & {\rm e^- + H_2 \rightarrow 2H + e^-}  & {\rm f(\sigma)} & [16]\\ 
            116 & {\rm e^- + H \rightarrow H^+ + 2e^-}  & {\rm f(\sigma)} & [13] \\ 
            117 & {\rm e^- + H^- \rightarrow H + 2e^-}  & {\rm f(\sigma)} & [17]\\ 
            118 & {\rm e^- + H_2^+ \rightarrow H^+ + H + e^-}  & {\rm f(\sigma)}& [18]\\
            119 & {\rm e^- + H_2^+ \rightarrow H^+ + H^-}  & {\rm f(\sigma)}& [19] \\ 
            120 & {\rm e^- + H_3^+ \rightarrow H^+ + H_2 + e^-}  & {\rm f(\sigma)}& [20] \\ 
            121 & {\rm e^- + H_3^+ \rightarrow H^+ + 2H + e^-}  & {\rm f(\sigma)}& [19] \\ 
            122 & {\rm e^- + H_3^+ \rightarrow 3H}  & {\rm f(\sigma)} & [19] \\ 
            123 & {\rm e^- + H_3^+ \rightarrow H_2 + H}  & 9.75\times10^{-8}T_e^{-0.5} & [21]\\ 
            124 & {\rm e^- + H^+ \rightarrow H} & 2.62\times10^{-13}T_e^{-0.5}   & [21] \\ 
            125 & {\rm e^- + H_2^+ \rightarrow 2H}  & 5.66\times10^{-8}T_e^{-0.5} & [21]\\ 
            \noalign{\smallskip}
            \hline\hline
         \end{array}
    $
\end{table*}

\begin{table*}
    %\caption{The CHO reactions included in SPARCKS are displayed here along with their rate coefficients and source. The rate coefficients are in $cm^3/s$.}
    \label{rcoefftable3}
    $ 
         \begin{array}{lcrc}
            \hline\hline
            \noalign{\smallskip}
            \text{RNO} & \multicolumn{1}{c}{\text{Reaction}} & \multicolumn{1}{l}{\text{Rate Coefficient} [{\rm cm^3/s}]} & \multicolumn{1}{l}{\text{Source}} \\
            \noalign{\smallskip}
            \hline
            \noalign{\smallskip}
            126 & {\rm H + H_2 \rightarrow 3H}  & 4.67\times10^{-7}(T_g/298)^{-1}\exp{(-5.5\times10^{4}/T_g)} & [10] \\ 
            127 & {\rm CH_5^+ + H \rightarrow CH_4^+ + H_2} & 1.5\times10^{-10} & [10]  \\ 
            128 & {\rm H_3^+ + CH_4 \rightarrow CH_5^+ + H_2} & 2.4\times10^{-9} & [10]  \\ 
            129 & {\rm H_3^+ + CH_3 \rightarrow CH_4^+ + H_2} & 2.1\times10^{-9}  & [10] \\ 
            130 & {\rm H_3^+ + CH_2 \rightarrow CH_3^+ + H_2} & 1.7\times10^{-9} & [10] \\ 
            131 & {\rm H_3^+ + CH \rightarrow CH_2^+ + H_2} & 1.2\times10^{-9} & [10] \\ 
            132 & {\rm H_3^+ + C \rightarrow CH^+ + H_2} & 2\times10^{-9} & [10] \\ 
            133 & {\rm H_3^+ + H^- \rightarrow H_2 + H_2} & 2.3\times10^{-7} & [10]  \\ 
            134 & {\rm H_2^+ + CH_2 \rightarrow CH_3^+ + H} & 1\times10^{-9} & [10] \\ 
            135 & {\rm H_2^+ + CH_2 \rightarrow CH_2^+ + H_2} & 1\times10^{-9} & [10] \\ 
            136 & {\rm H_2^+ + CH \rightarrow CH_2^+ + H} & 7.1\times10^{-10}(T_g/300)^{(-0.5)} & [14]\\ 
            137 & {\rm H_2^+ + CH \rightarrow CH^+ + H_2} & 7.1\times10^{-10}(T_g/300)^{(-0.5)} & [14]\\ 
            138 & {\rm H_2^+ + C \rightarrow CH^+ + H} & 2.4\times10^{-9} & [10] \\ 
            139 & {\rm H_2^+ + H_2 \rightarrow H_3^+ + H} & 2.08\times10^{-9} & [10] \\ 
            140 & {\rm H_2^+ + H \rightarrow H^+ + H_2} & 6.4\times10^{-10} & [10] \\ 
            141 & {\rm H^+ + CH_3 \rightarrow CH_3^+ + H} & 3.4\times10^{-9} & [10] \\ 
            142 & {\rm H^+ + CH_2 \rightarrow CH_2^+ + H} & 1.4\times10^{-9} & [10] \\ 
            143 & {\rm H^+ + CH_2 \rightarrow CH^+ + H_2} & 1.4\times10^{-9} & [10] \\ 
            144 & {\rm H^+ + CH \rightarrow CH^+ + H} & 1.9\times10^{-9} & [14] \\ 
            145 & {\rm H^- + CH_2 \rightarrow CH_3 + e^-} & 1\times10^{-9} & [10] \\ 
            146 & {\rm H^- + CH \rightarrow CH_2 + e^-} & 1\times10^{-10} & [10] \\ 
            147 & {\rm H^- + C \rightarrow CH + e^-} & 1\times10^{-9} & [10] \\ 
            148 & {\rm H^- + H \rightarrow H_2 + e^-} & 1.3\times10^{-9} & [10] \\ 
            149 & {\rm C^+ + H^- \rightarrow C + H} & 2.3\times10^{-7} & [10]\\ 
            150 & {\rm H_2^+ + H^- \rightarrow H_2 + H} & 2.3\times10^{-7} & [10]\\ 
            151 & {\rm H^+ + H^- \rightarrow H + H} &  2.3\times10^{-7} & [10]\\ 
            152 & {\rm H^- + H_3^+ \rightarrow H_2 + 2H} &  1\times10^{-7} & [10]\\ 
            153 & {\rm CH_2 + H_2 \rightarrow CH_3 + H} & 5\times10^{-15} & [10]\\ 
            154 & {\rm CH_3 + OH \rightarrow CH_4 + O} & 3.22\times10^{-14}(T_g/298)^{2.2}\exp{(-18.62/(RT_g))} & [10] \\ 
            155 & {\rm e^- + CO \rightarrow C^+ + O + 2e^-} & {\rm f(\sigma)} & [5]\\ 
            156 & {\rm e^- + CO \rightarrow C + O^+ + 2e^-} & {\rm f(\sigma)} & [5]\\ 
            157 & {\rm e^- + CO \rightarrow C + O + e^-} & {\rm f(\sigma)} & [5]\\ 
            158 & {\rm e^- + CO_2 \rightarrow C^+ + O_2} + 2e^- & {\rm f(\sigma)} & [4]\\ 
            159 & {\rm e^- + H_2O \rightarrow H_2 + O^-} & {\rm f(\sigma)} & [22]\\ 
            160 & {\rm e^- + H_2O \rightarrow H + OH^-} & {\rm f(\sigma)} & [22]\\ 
            161 & {\rm e^- + H_2O \rightarrow H + OH + e^-} & {\rm f(\sigma)} & [22]\\ 
            162 & {\rm e^- + H_2O \rightarrow H_2 + O + e^-} & {\rm f(\sigma)} & [22] \\ 
            163 & {\rm e^- + H_2O \rightarrow H_2O^+ + 2e^-} & {\rm f(\sigma)} & [22]\\ 
            164 & {\rm e^- + OH \rightarrow OH^+ + 2e^-} & 2.0\times10^{-10}T_e^{1.78}\exp{(-13.8/T_e)} & [23]\\ 
            165 & {\rm e^- + OH \rightarrow O + H + e^-} & 2.08\times10^{-7}T_g^{-0.76}  \exp{(-6.9/T_e)} &[23] \\ 
            166 & {\rm e^- + OH^- \rightarrow OH + 2e^-} & 9.67\times10^{-6}T_e^{-1.89}\exp{-12.1/T_e} &[23]\\ 
            167 & {\rm e^- + OH^- \rightarrow O + H + 2e^-} & 1.95\times10^{-8} & [10] \\ 
            168 & {\rm e^- + O(3P) \rightarrow O^+ + 2e^-}     & {\rm f(\sigma)} & [13]\\ 
            169 & {\rm e^- + O(3P) \rightarrow O(1D) + e^-}   & {\rm f(\sigma)} & [13]\\ 
            170 & {\rm e^- + H_2O^+ \rightarrow OH + H}        & 0.22.6\times10^{-7}   & [24] [25]\\ 
            171 & {\rm e^- + H_2O^+ \rightarrow O + H_2}        & 0.092.6\times10^{-7}   & [24] [25]   \\ 
            172 & {\rm e^- + H_2O^+ \rightarrow O + 2H}        & 0.71e02.6\times10^{-7}  & [24] [25] \\ 
            173 & {\rm e^- + OH^+ \rightarrow O + H}          & 2\times10^{-7}(T_e/300)^{-0.5} & [23]\\ 
            174 & {\rm e^- + CO_2^+ \rightarrow C + O_2}        & 1.7\times10^{-6}(T_e/300)^{-0.4} & [7] \\ 
            175 & {\rm e^- + O_2^+ \rightarrow O + O}          & 1.49\times10^{-7}(300/(T_e))^{0.7} & [6]\\ 
            176 & {\rm CH_4 + O \rightarrow CH_3 + OH}        & 8.32\times10^{-12}(T_g/298)^{1.56}\exp{(-35.503/(RT_g))} & [10] \\ 
            177 & {\rm CH_3 + O \rightarrow CO + H_2 + H}     & 2.8\times10^{-11} & [10] \\ 
            178 & {\rm CH_2 + O \rightarrow CO + H_2}         & 5.53\times10^{-11} & [10] \\
            179 & {\rm CH_2 + O \rightarrow CO + 2H}         & 8.29\times10^{-11} & [10] \\ 
            180 & {\rm CH_2 + O_2 \rightarrow CO_2 + H_2}       & 2.92\times10^{-11}(T_g/300)^{-3.3}\exp{(-1443/T_g)} & [26] \\ 
            181 & {\rm CH_2 + O_2 \rightarrow CO + H_2O}       & 2.48\times10^{-10}(T_g/300)^{-3.3}\exp{(-1443/T_g)} & [26]\\ 
            182 & {\rm CH + O \rightarrow CO + H}           & 6.59\times10^{-11} & [10] \\ 
            183 & {\rm CH + O_2 \rightarrow CO_2 + H}         & 1.2\times10^{-11} & [10]\\ 
            184 & {\rm CH + O_2 \rightarrow CO + H + O}      & 1.2\times10^{-11}  & [10]\\ 
            185 & {\rm C + O_2 \rightarrow CO + O}           & 1.1\times10^{-10}\exp{(-320/T_g)} & [11]\\ 
            186 & {\rm H_2 + O \rightarrow OH + H}           & 2.21\times10^{-12}(T_g/2.98)^{2.0}\exp{(-31.595/(RT_g))} & [10] \\ 
            187 & {\rm H + O_2 \rightarrow OH + O}           & 1.62\times10^{-10}\exp{(-62.109/(RT_g))} & [10] \\ 
            188 & {\rm OH + O \rightarrow H + O_2}           & 4.55\times10^{-12}(T_g/298)^{0.4}\exp{(-3.097/(RT_g))} & [10]  \\ 
            189 & {\rm CH_3 + OH \rightarrow CH_2 + H_2O}      & 1.2\times10^{-10}\exp{(-11.64/(RT_g))} & [10]  \\ 
            
            \noalign{\smallskip}
            \hline\hline
         \end{array}
    $
\end{table*}

\begin{table*}
    %\caption{The CHO reactions included in SPARCKS are displayed here along with their rate coefficients and source. The rate coefficients are in $cm^3/s$.}
    \label{rcoefftable4}
    $ 
         \begin{array}{lcrc}
            \hline\hline
            \noalign{\smallskip}
            \text{RNO} & \multicolumn{1}{c}{\text{Reaction}} & \multicolumn{1}{l}{\text{Rate Coefficient} [{\rm cm^3/s}]} & \multicolumn{1}{l}{\text{Source}} \\
            \noalign{\smallskip}
            \hline
            \noalign{\smallskip}
            190 & {\rm H_2 + OH \rightarrow H + H_2O}         & 2.08\times10^{-12}(T_g/300)^{1.52}\exp{(-1740/T_g)} & [14] \\
            191 & {\rm H + H_2O \rightarrow H_2 + OH}         & 6.82\times10^{-12}(T_g/298)^{1.6}\exp{(-80.82/(RT_g))} & [10] \\ 
            192 & {\rm H + CO_2 \rightarrow CO + OH}         & 2.51\times10^{-10}\exp{(-111.0/(RT_g))} & [10] \\ 
            193 & {\rm O + H_2O \rightarrow 2OH}             & 1.85\times10^{-11}(T_g/300)^{0.95}\exp{(-8571/T_g)} & [14] \\ 
            194 & {\rm 2OH \rightarrow O + H_2O}             & 1.26\times10^{-11}\exp{(-4.182/(RT_g))} & [10] \\ 
            195 & {\rm CO_2 + C \rightarrow CO + O_2}         & 2.8\times10^{-11}\exp{(-26500/T_g)} &[10] \\ 
            196 & {\rm CO_2 + C \rightarrow 2CO}             & 1.\times10^{-15} & [10] \\ 
            197 & {\rm O_2 + CO \rightarrow CO_2 + O}         & 4.2\times10^{-12}\exp{(-24000/T_g)} & [10] \\ 
            198 & {\rm O_2 + O_2 \rightarrow O_3 + O}          & 2.1\times10^{-11}\exp{(-498000/T_g)} & [10] \\ 
            199 & {\rm O_2 + O_2 \rightarrow O + O + O_2}      & 3.6\times10^{-8}(1.-\exp{(-2240/T_g))}\exp{(-59380/T_g)} & [10] \\ 
            200 & {\rm O_2 + O \rightarrow O + O + O} & 1.28\times10^{-7}(1-\exp{(-2240/T_g))}\exp{(-59380/T_g)} & [10] \\ 
            201 & {\rm CH_2 + O_2 \rightarrow CO_2 + 2H}       & 1.08\times10^{-11}\exp{(-758/T_g)} & [10] \\ 
            202 & {\rm CH_2 + O_2 \rightarrow CO + OH + H}    & 1.08\times10^{-11}\exp{(-758/T_g)} & [10] \\ 
            203 & {\rm CH + O \rightarrow OH + C}           & 2.52\times10^{-11}\exp{(-2380/T_g)} & [10] \\ 
            204 & {\rm O^*+ CH_4 \rightarrow CH_3 + OH}       & 3.11\times10^{-10} &[10] \\ 
            205 & {\rm O^*+ CO \rightarrow CO_2}             & 8.00\times10^{-11} &[10] \\ 
            206 & {\rm O^*+ CO \rightarrow CO + O}          & 5.00\times10^{-11} &[10] \\ 
            207 & {\rm O^*+ H_2O \rightarrow OH + OH}        & 2.2\times10^{-10} &[10] \\ 
            208 & {\rm O^+ + CH_4 \rightarrow CH_4^+ + O}       & 8.9\times10^{-10} &[10] \\ 
            209 & {\rm O^+ + CH_4 \rightarrow CH_3^+ + OH}      & 1.1\times10^{-10} &[10] \\ 
            210 & {\rm O^+ + CH_2 \rightarrow CH_2^+ + O}       & 9.7\times10^{-10} &[10] \\ 
            211 & {\rm O^+ + CH \rightarrow CH^+ + O}         & 3.5\times10^{-10}(T_g/300)^{-0.5} &[14] \\ 
            212 & {\rm O^+ + CH \rightarrow CO^+ + H}         & 3.5\times10^{-10}(T_g/300)^{-0.5} &[14]\\ 
            213 & {\rm O^+ + H_2 \rightarrow OH^+ + H}         & 1.35\times10^{-9} &[14]  \\ 
            214 & {\rm O^+ + H \rightarrow H^+ + O}           & 5.66\times10^{-10}(T_g/300)^{0.36}\exp{(8.6/T_g)} & [14]\\ 
            215 & {\rm O^+ + OH \rightarrow OH^+ + O}         & 3.6\times10^{-10}(T_g/300)^{-0.5} &  [14] \\ 
            216 & {\rm O^+ + OH \rightarrow O_2^+ + H}         & 3.6\times10^{-10}(T_g/300)^{-0.5} &[14] \\ 
            217 & {\rm O^+ + H_2O \rightarrow H_2O^+ + O}       & 3.2\times10^{-9}(T_g/300)^{-0.5} &[14] \\ 
            218 & {\rm O^+ + CO_2 \rightarrow O_2^+ + CO}       & 9.4\times10^{-10} &[10] \\ 
            219 & {\rm CH + O_2 \rightarrow CO + OH}         & 8.0\times10^{-12}& [10] \\ 
            220 & {\rm CH_4^+ + O \rightarrow CH_3^+ + OH}      & 1.\times10^{-9} &[10] \\ 
            221 & {\rm CH_4^+ + O_2 \rightarrow O_2^+ + CH_4}     & 3.9\times10^{-10} &[10] \\ 
            222 & {\rm CH^+ + O \rightarrow CO^+ + H}         & 3.5\times10^{-10} &[10] \\ 
            223 & {\rm CH^+ + O_2 \rightarrow CO^+ + OH}       & 1.\times10^{-11} &[10] \\ 
            224 & {\rm CH^+ + OH \rightarrow CO^+ + H_2}       & 7.5\times10^{-10}(T_g/300)^{-0.5} & [14] \\ 
            225 & {\rm C^+ + O_2 \rightarrow O^+ + CO}         & 4.54\times10^{-10} & [14] \\ 
            226 & {\rm C^+ + O_2 \rightarrow CO^+ + O}         & 3.42\times10^{-10} & [14] \\ 
            227 & {\rm C^+ + OH \rightarrow CO^+ + H}         & 7.7\times10^{-10}(T_g/300)^{-0.5} & [14]\\ 
            228 & {\rm C^+ + CO_2 \rightarrow CO^+ + CO}       &1.1\times10^{-9} & [10] \\ 
            229 & {\rm H^+ + O \rightarrow O^+ + H}           & 6.86\times10^{-10}(T_g/300)^{0.26}\exp{(-224.3/T_g)} & [14]\\ 
            230 & {\rm H^+ + O_2 \rightarrow O_2^+ + H}         & 2.\times10^{-9} & [10] \\ 
            231 & {\rm H^+ + OH \rightarrow OH^+ + H}         & 2.1\times10^{-9}(T_g/300)^{-0.5} & [14]\\ 
            232 & {\rm H^+ + H_2O \rightarrow H_2O^+ + H}       & 8.2\times10^{-9} & [15]  \\ 
            233 & {\rm CH_3 + H_2O \rightarrow CH_4 + OH}      & 1.2\times10^{-14}(T_g/298)^{2.9}\exp{(-62.19/(RT_g))} & [10] \\ 
            234 & {\rm O_2^+ + CH_2 \rightarrow CH_2^+ + O_2}     & 4.3\times10^{-10} &[10] \\ 
            235 & {\rm O_2^+ + CH \rightarrow CH^+ + O_2}       & 3.1\times10^{-10} & [10] \\ 
            236 & {\rm O_2^+ + C \rightarrow CO^+ + O}         & 5.2\times10^{-11} & [10] \\ 
            237 & {\rm O_2^+ + C \rightarrow C^+ + O_2}         & 5.2\times10^{-11}  &[10] \\ 
            238 & {\rm H_2 + e^- \rightarrow H^- + H}          & 2.3\times10^{-13}T_e^{(1/2)}(1.+4.6\times10^{-5}T_e)\exp{(-4.3\times10^{-4}/T_e)} & [7]  \\ 
            239 & {\rm CO + OH \rightarrow H + CO_2}         & 5.4\times10^{-14}(T_g/298)^{1.5}\exp{(2.08/(RT_g))} & [10] \\ 
            240 & {\rm CH_4 + OH \rightarrow CH_3 + H_2O}      & 1.36\times10^{-13}(T_g/298)^{3.04}\exp{(-7.649/(RT_g))} & [10]\\ 
            \noalign{\smallskip}
            \hline\hline
            \noalign{\smallskip}
            \multicolumn{4}{@{}l}{\text{[1] \cite{ionin2007physics} [2] \cite{gudhmundsson2004critical} [3] \cite{1997nahar} [4] \cite{itikawa2002cross} [5] \cite{itikawa2015cross} } }\\
            \multicolumn{4}{@{}l}{\text{[6] \cite{hokazono1991theoretical} [7] \cite{beuthe1997chemical} [8] \cite{bouwman2021neutral} [9] \cite{song2015cross}  } }\\
            \multicolumn{4}{@{}l}{\text{[10] \cite{wang2018} [11] \citep{baulch2005evaluated} [12] \cite{janev2002collision} [13] \cite{morgankinema} [14] \cite{millar2024umist}} } \\
            \multicolumn{4}{@{}l}{\text{[15] \cite{anicich1986survey} [16] \cite{yoon2008cross} [17] \cite{janev2003collision} [18] \cite{1987janev}} }\\
            \multicolumn{4}{@{}l}{\text{[19] \cite{tawara1990cross} [20] \cite{chan1983model} [21] \cite{salabas2004systematic} [22] \cite{itikawa2005cross}} }\\
            \multicolumn{4}{@{}l}{\text{[23] \cite{liu2010global} [24] \cite{jensen1999dissociative} [25] \cite{rosen2000recombination} [26] \cite{dombrowsky1992investigation}}}\\
            \noalign{\smallskip}
         \end{array}
    $
\end{table*}

%%%%%%%%%%%%%%%%%%%%%%%%%%%%%%%%%%%%%%%%%%%%%%%%%%

% Don't change these lines
\bsp	% typesetting comment
\label{lastpage}
\end{document}